\documentclass[11pt,a4paper]{article}
\usepackage{jcappub}

\usepackage{hyperref}
\usepackage{braket}
\usepackage{amsmath}
\usepackage{amsfonts}
\usepackage[dvipsnames]{xcolor}
\usepackage{graphicx}
\usepackage{caption}
\usepackage{subcaption}

\begin{document}

\title{Gauged Quintessence}
\author[a]{Kunio Kaneta,}
\author[b]{Hye-Sung Lee,}
\author[b]{Jiheon Lee,}
\author[b]{Jaeok Yi}
\affiliation[a]{Department of Mathematics, Tokyo Woman’s Christian University, Tokyo 167-8585, Japan}
\affiliation[b]{Department of Physics, KAIST, Daejeon 34141, Korea}
\abstract{
\noindent
Despite its dominance in the present universe's energy budget, dark energy is the least understood component in the universe.
Although there is a popular model for the dynamical dark energy, the quintessence scalar, the investigation is limited because of its highly elusive character.
We present a model where the quintessence is gauged by an Abelian gauge symmetry.
The quintessence is promoted to be a complex scalar whose real part is the dark energy field while the imaginary part is the longitudinal component of a new gauge boson.
It brings interesting characters to dark energy physics.
We study the general features of the model, including how the quintessence behavior is affected and how the solicited dark energy properties constrain its gauge interaction.
We also note that while the uncoupled quintessence models are suffered greatly from the Hubble tension, it can be alleviated if the quintessence is under the gauge symmetry.
}
\maketitle
\flushbottom

\section{Introduction}
Dark energy is very elusive because it started to dominate the universe's energy density only recently ($z\lesssim1$).
While simple and quite successful, the dark energy in the $\Lambda$-CDM model is treated as a non-dynamical cosmological constant ($\Lambda$) of the general relativity, which is only determined by fitting the model with data.
(The CDM stands for cold dark matter.) Taking the cosmological constant for the dark energy is not satisfying for many reasons including the lack of reason why we have a similar density of dark matter and dark energy in the present universe (cosmological coincidence problem). The distance conjecture on the swampland also disapproves the stable de Sitter vacuum \cite{Ooguri:2018wrx,Garg:2018reu}, and favors a dynamically varying vacuum over a constant one.

Quintessence is a popular model suggested by Ratra and Peebles that identifies the dark energy as a scalar field, which slowly rolls down a potential \cite{PhysRevD.37.3406}.
The dynamical feature of the quintessence models allows them to address the cosmological coincidence problem \cite{Zlatev:1998tr}. (See Refs.~\cite{Martin:2008qp,Tsujikawa:2013fta} for some reviews on the quintessence.) 

Established evidence of dark matter, the other component of the unknown part of the universe, is also based only on the gravitational effects.
However, direct/indirect dark matter searches are based on the assumption that dark matter has a sizable interaction other than gravity with the standard model (SM) particles.
It may be the SM interaction or a new one such as a new gauge symmetry \cite{Silveira:1985rk,McDonald:1993ex,Burgess:2000yq,Djouadi:2011aa,Djouadi:2012zc,Mambrini:2011ik,Alves:2013tqa,Lebedev:2014bba,Arcadi:2013qia,Arcadi:2014lta,Ellis:2017ndg,Escudero:2016gzx,Hall:2009bx,Bhattacharyya:2018evo,Kaneta:2016vkq,Kaneta:2016wvf,Anastasopoulos:2020gbu,Brax:2020gqg,Brax:2021gpe,Kaneta:2021pyx,Chowdhury:2018tzw,Kamada:2018zxi}.

Considering the myriad of global activities in the dark matter searches based on the dark matter interaction, it may be worth investigating if the quintessence may have interaction and what constraints are expected in such models.
The new interaction might open a new direction for studying dark energy.\footnote{In fact, it was already discussed on general grounds that the scattering of dark energy and baryon could affect the cosmic evolution \cite{Vagnozzi:2019kvw,Ferlito:2022mok} and the screened coupling of dark energy to photon may explain the XENON1T anomaly \cite{Vagnozzi:2021quy}.}
It would be important to build a specific and realistic model of a new interaction to see what constraints generally arise and what predictions are naturally given.

In this paper, we present a dark energy model ``gauged quintessence'',\footnote{We note the ``gauge quintessence'' model \cite{Pilo_2003}, which takes a vector boson as a dark energy field. Our ``gauged quintessence'' model takes a scalar as a dark energy field charged by a gauge symmetry, and the gauge boson is not the quintessence.} in which the quintessence scalar field is the radial part of a complex scalar charged by an Abelian gauge symmetry.
The scalar boson and the gauge boson of this model are of tiny scale, and the original quintessence scalar is restored in the limit the gauge coupling vanishes.
Dark energy physics is quite sensitive to the interplay between the quintessence scalar and the dark gauge boson, and we exploit it to constrain the model.
We also address the Hubble tension issue \cite{Verde:2019ivm} with this quintessence model under a gauge symmetry. The uncoupled quintessence model is known to be agonized by the Hubble tension much greater than the $\Lambda$-CDM model \cite{Banerjee:2020xcn,Lee:2022cyh}. We discuss how the gauge symmetry can alleviate this drawback.

Taking quintessence as a part of a complex scalar field is not a new idea \cite{Gu:2001tr,Boyle:2001du,Brisudova:2001wb,Brisudova:2001ur,Li:2001xaa,Mainini:2004he,Frieman:1995pm,Kim:1998kx,Choi:1999xn,Kim:2002tq,Hill:2002kq}.
The complex scalar under a global symmetry was studied in Refs.~\cite{Frieman:1995pm,Kim:1998kx,Choi:1999xn,Carroll:1998zi,Kim:2002tq,Hill:2002kq}.
As a matter of fact, it was pursued with a gauge symmetry, too. The authors of Refs.~\cite{Brisudova:2001wb,Brisudova:2001ur} studied the homogeneous scalar field carrying a $U(1)$ charge to see if a long-range repulsive force can explain the late time accelerating expansion of the universe. The gauge symmetry in this scenario needs to be explicitly broken to have a nonzero homogeneous charge density \cite{Brisudova:2001wb}.
They considered homogeneous and isotropic gauge field configuration (insisting their gauge field, $\mathbb{X}_\mu$, should satisfy $\partial_{i}\mathbb{X}_0=0$ and $\vec{\mathbb{X}}=0$), which results that the gauge field is not a dynamical field of its equation of motion.

In contrast, our model preserves the gauge symmetry and there is no explicit symmetry breaking.
The gauge boson mass is given by the value of the quintessence scalar, which is basically the spontaneous breaking of the $U(1)$ gauge symmetry.
Furthermore, in our model, the gauge field maintains dynamical degrees of freedom.
Therefore, we can take into account the dark gauge boson quanta or its zero momentum condensates.
The charge density is zero, yet the gauge symmetry alters the quintessence scalar potential by the dynamical gauge field.

The basic structure of the model is described in Sec.~\ref{sec:model}.
In Sec.~\ref{sec:quantum correction}, we calculate quantum corrections and argue why a tiny gauge coupling is demanded to protect the late time slow-roll of the scalar field, which is essential to produce the dark energy-like behavior.
We discuss the general behavior of the quintessence scalar and the dark gauge boson in Sec.~\ref{sec:Quintessence dynamics}.
In Sec.~\ref{sec:Gauged quintessence on the $H_0$ tension}, we discuss the impact of the gauged quintessence on the Hubble tension.
We discuss the constraints on the model in Sec.~\ref{sec:cons_on_gauge} before the summary and outlook in Sec.~\ref{sec:Summary}.
In App.~\ref{sec:Slow roll condition}, we go over basic slow-roll conditions for the quintessence field.
In App.~\ref{sec:Redshift of the gauge potential}, we discuss the evolution of the gauge symmetry induced potential and derive the Boltzmann equations for mass varying particles.
In App.~\ref{sec:Ultra-light dark gauge boson}, we obtain the potential for a coherent dark gauge boson.

\section{Model}\label{sec:model}
The model consists of a complex scalar ($\Phi$) and a $U(1)_\text{Dark}$ gauge boson field ($\mathbb{X}^{\mu}$). The complex scalar field has a radial ($\phi$) and angular ($\eta$) degrees of freedom,
\begin{equation}
\Phi=\frac{1}{\sqrt{2}}\phi\, e^{i\eta} \, ,
\end{equation}
which is charged by a local $U(1)_\text{Dark}$ symmetry and is singlet under the SM gauge symmetries. We will treat $\phi$ as a cosmologically running field, which is homogeneous and isotropic,\footnote{This condition may be relaxed. For instance, see Refs.~\cite{Motoa-Manzano:2020mwe,Orjuela-Quintana:2020klr,Guarnizo:2020pkj,Orjuela-Quintana:2021zoe} for anisotropic dark energy models.} and take it as a quintessence dark energy field. (We assume $\phi>0$ throughout this paper.) The nonzero value of $\phi$ gives the mass to the dark gauge boson, and the $\eta$ becomes the longitudinal component of the dark gauge boson $\mathbb{X}^\mu$ of the $U(1)_\text{Dark}$. 
We call our model {\em gauged quintessence} model.

The minimal gauge-invariant action containing the $\Phi$ and the gauge boson in the isotropic, homogeneous, and flat universe, i.e., $g_{\mu\nu}=\text{Diag}\{-1,a(t)^2,a(t)^2,a(t)^2\}$, is
\begin{equation}
S = \int d^4x \; \sqrt{-g}\Big[\frac{1}{2} m_{Pl}^2 R -|D_{\mu}\Phi|^2-V_0(\Phi)-\frac{1}{4}\mathbb{X}_{\mu\nu}\mathbb{X}^{\mu\nu} \Big] \, ,
\label{eq:action}
\end{equation}
where $m_{Pl}=M_{Pl}/\sqrt{8\pi}\approx 2.4\times10^{18}$ GeV is the reduced Planck mass with $M_{Pl}\approx 1.2\times 10^{19}$ GeV being the Planck mass, $R$ is the Ricci scalar representing the curvature, $D_\mu \equiv \partial_\mu + i g_X \mathbb{X}_\mu$ with a $U(1)_\text{Dark}$ gauge coupling constant $g_X$, $\mathbb{X}_{\mu\nu} \equiv \partial_\mu \mathbb{X}_\nu-\partial_\nu \mathbb{X}_\mu$, and $V_0$ is the potential for the scalar field only.
In the unitary gauge where
\begin{equation}
    \eta= 0 , \quad X_{\mu} = \mathbb{X}_{\mu} +\frac{1}{g_X}\partial_{\mu}\eta \, ,
\end{equation}
the imaginary part of the $\Phi$ is absorbed into the dark gauge boson field, and the action can be written as
\begin{equation}
    S= \int d^4x \; \sqrt{-g}\Big[\frac{1}{2}m_{Pl}^2 R -\frac{1}{2}(\partial_{\mu}\phi)^2 -\frac{1}{4}X_{\mu\nu}X^{\mu\nu}-V_0(\phi) -\frac{1}{2}g_X^2\phi^2X_{\mu}X^{\mu} \Big] \, .
\end{equation}

The last term in the action is the interaction between the quintessence scalar and the dark gauge boson, which contributes to both the dark gauge boson and the quintessence masses. This term contributes to the quintessence potential, and we call this term the gauge potential ($V_{\text{gauge}}$),
\begin{equation}
    V_{\text{gauge}}= \frac{1}{2}g_X^2\phi^2X_{\mu}X^{\mu}.
\end{equation}
Then the tree-level masses are
\begin{equation}
m_{\phi}^2|_0 = \frac{\partial^2 V_0}{\partial \phi^2} + \frac{\partial^2 V_{\text{gauge}}}{\partial \phi^2} \, , \quad m_X^2|_0 = g_X^2 \phi^2 \, .
\label{eq:treemass}
\end{equation}
Here, the symbol $|_0$ indicates that the corresponding mass is evaluated by using a value of $\phi(t)$ determined by the tree-level potential for $\phi$, namely, $V_0(\phi)$.
The quintessence mass and the dark gauge boson mass change during cosmic evolution.

To discuss the background evolution, one should know how the $\phi$ and $X$ evolve during the cosmic evolution. The equations of motion for the $\phi$ and $X^{\mu}$ with the tree-level potential are given by
\begin{equation}
\begin{split}  
\ddot{\phi}  +3 H \dot{\phi}+\frac{\partial V_0}{\partial \phi}+g_X^2 X_{\mu}X^{\mu} \phi&=0\, ,\\    \partial_\mu X^{\mu\nu}+3HX^{0\nu}-g_X^2\phi^2X^\nu& = 0 \,.
\end{split}
\label{eq:eom}
\end{equation}
Thus the evolution of the quintessence is affected by the background dark gauge boson field.

The energy density ($\rho_{\phi+X}$) and pressure ($p_{\phi+X}^{i}$) of the quintessence scalar and the dark gauge boson are obtained from the $00$ and $ii$ component of the Hilbert stress-energy tensor,
\begin{multline}
T_{\mu\nu}=(\partial_{\mu}\phi)(\partial_{\nu}\phi) -\frac{1}{2} g_{\mu\nu}(\partial_{\alpha}\phi)(\partial^{\alpha}\phi)-g_{\mu\nu}V_0(\phi)\\ -\frac{1}{2}g_{\mu\nu} g_X^2 \phi^2 X_{\alpha}X^{\alpha}+g_X^2\phi^2 X_{\mu}X_{\nu}+ X_{\mu\alpha} X_{\nu}^{\alpha}-\frac{g_{\mu\nu}}{4} X_{\alpha\beta} X^{\alpha\beta} .
\end{multline}
In the following, we divide the stress-energy tensor into two parts, one corresponding to the quintessence and the rest corresponding to the dark gauge boson. We define the quintessence energy density ($\rho_{\phi}$) and ($p_{\phi}$) as the same as the usual definition of the scalar field energy density and pressure:
\begin{equation}
\begin{split}
    \rho_{\phi} &= \frac{1}{2}\dot{\phi}^2 +V_0(\phi) \, , \\
    p_{\phi} &= \frac{1}{2}\dot{\phi}^2 -V_0(\phi) \, .
\end{split}
\label{eq:phidensandP}
\end{equation}
Also, the energy density and the pressure of the dark gauge boson is $\rho_X=\rho_{\phi+X}-\rho_{\phi}$, and $p_X=p_{\phi+X}-p_{\phi}$.

Even if the $V_{\text{gauge}}$ is not a part of the $\rho_{\phi}$, this term still affects the motion of the quintessence scalar. Therefore, the scaling of the quintessence energy density by the scale factor is not given by the typical definition of the equation of state,
\begin{equation}
    w_0= \frac{p_{\phi}}{\rho_{\phi}} \, .
\end{equation} 
Instead, one can define the effective equation of state, $w_{\text{eff}}$ \cite{Brookfield:2005bz},
\begin{equation}
 \dot{\rho}_{\phi} +3H(1+w_{\text{eff}})\rho_{\phi}=0 \, , 
 \label{eq:findeos}
\end{equation}
where the explicit form of the $w_{\text{eff}}$ comes from Eqs.~\eqref{eq:eom} and \eqref{eq:findeos},
\begin{equation}
    w_{\text{eff}}= w_0 + \frac{1}{3H\rho_{\phi}}\frac{\partial V_{\text{gauge}}}{\partial \phi}\dot{\phi} \, .
    \label{eq:weff}
\end{equation}
When the $\rho_{\phi}\gg \rho_{X}$, the quintessence scalar can explain the dark energy if $w_{\text{eff}}<-1/3$.

For the scalar potential $V_0$ in our analysis, we take the inverse power potential suggested in Ratra-Peeble's work \cite{PhysRevD.37.3406}
\begin{equation}
V_0 (\phi) = \frac{M^{\alpha+4}}{\phi^{\alpha}} \, ,
\label{eq:RPpotential}
\end{equation}
where $\alpha>0$ and the mass scale $M$ will be specified later.

\section{Quantum corrections}\label{sec:quantum correction}
Calculating quantum corrections to the quintessence potential, using the functional method \cite{PhysRevD.7.1888,Jackiw:1974cv,Doran:2002bc}, is in order.
The dark energy behavior of the quintessence scalar depends on the potential, and we have to ensure the right property is preserved even at the quantum corrected level.

\begin{figure}[t]
    \centering
\begin{subfigure}{0.75\textwidth}
    \includegraphics[width=0.99\linewidth]{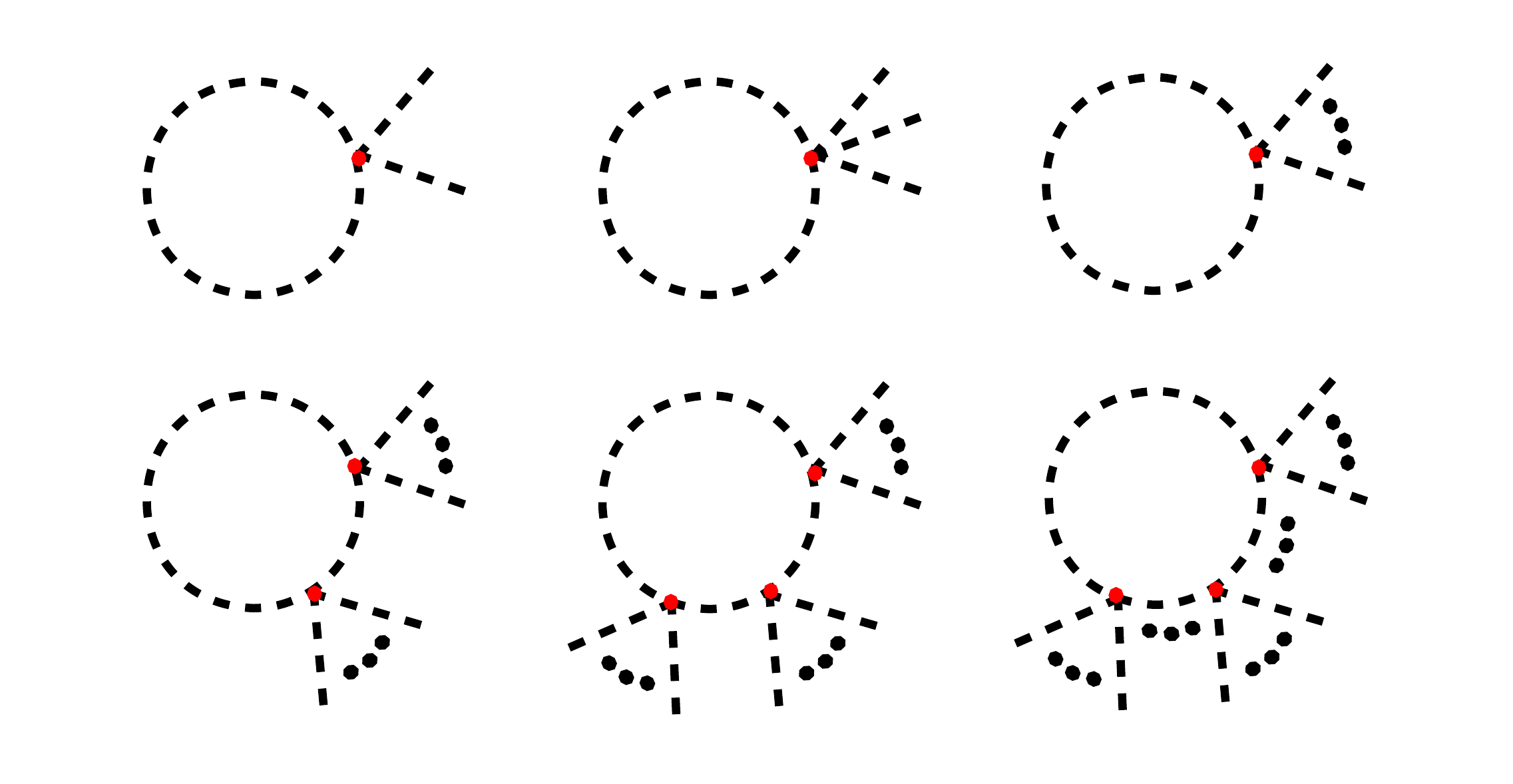}
    \caption{}
\end{subfigure}
\begin{subfigure}{0.75\textwidth}
    \includegraphics[width=0.99\linewidth]{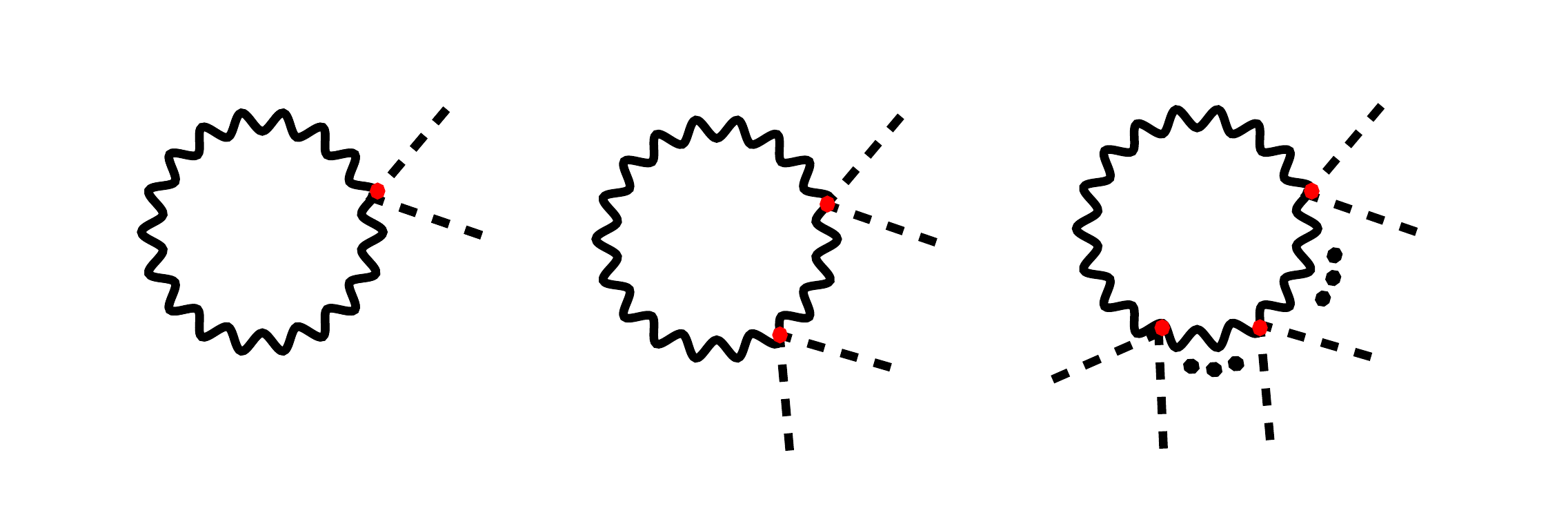}
    \caption{}
    \end{subfigure}
    \caption{1-loop Feynman diagrams in the gauged quintessence model. The dashed are for the quintessence, and the waves are for the dark gauge boson. The diagrams in (a) correspond to the two terms involving $V_0^{\prime\prime}$ in Eq.~\eqref{eq:QV}, and those in (b) correspond to the last term in Eq.~\eqref{eq:QV}.}
    \label{fig:feynman}
\end{figure}

The quantum effective potential of the scalar field ($V_{\text{eff}}$) is obtained from the saddle point approximation 
 \begin{equation}
   V_{\text{eff}} = -\frac{1}{VT} \Gamma[\phi_\text{cl}] =-\frac{1}{VT}\left( \int \frac{d^4x}{(2\pi)^4}\, \mathcal{L}[\phi_
     \text{cl}]+ \frac{i}{2}\, \ln\, \text{det}
     \Big(\frac{\delta^2\mathcal{L}}{\delta\phi\delta\phi}\Big) + \cdots \right) ,
\end{equation}
where $VT$ is a spacetime volume. The $\phi_\text{cl}$ is the vacuum expectation value of $\phi$. We assume the homogeneous vacuum; thus, $\phi_{\text{cl}}$ only depends on time.
The determinant is evaluated over all the internal degrees of freedom and gives the 1-loop quantum correction. This is equivalent to the sum of all the 1-loop diagrams in Fig.~\ref{fig:feynman}. 

Then the $V_{\text{eff}}$ with a cutoff $\Lambda$ is given as 
\begin{equation}
    V_{\text{eff}}= V_0 +\frac{1}{2}g_X^2 X_{\mu}X^{\mu}\phi^2 + \frac{\Lambda^2}{32\pi^2}V_0^{\prime\prime} + \frac{(V_0^{\prime\prime})^2}{64\pi^2} \left(\ln\frac{V_0^{\prime\prime}}{\Lambda^2}-\frac{3}{2}\right) +\frac{3(m_{\text{X}}^2|_0)^2}{64\pi^2} \left(\ln\frac{m_{\text{X}}^2|_0}{\Lambda^2}-\frac{5}{6} \right) ,
    \label{eq:QV}
\end{equation} 
where the $\prime$ denotes the partial derivative with respect to the $\phi$. The quadratic divergence term from the gauge boson loop can be absorbed to the counter term quadratic in $\phi$. For the effective potential to have a physical meaning, it should be gauge-independent. With our Lagrangian in unitary gauge, the gauge dependence of the effective potential is removed \cite{Dolan:1974gu,Nielsen:1975fs}. 

Now one can obtain the $\phi$ mass with the leading order correction from the second derivative of $V_\text{eff}$ as
\begin{equation}
    m_{\phi}^2 = V_0^{\prime\prime} + g_X^2  X_{\mu}X^{\mu} +\frac{\Lambda^2}{32\pi^2} V_0^{\prime\prime\prime\prime}+ \frac{V_0^{\prime\prime}V_0^{\prime\prime\prime\prime}}{32\pi^2}\left(\ln\frac{V_0^{\prime\prime}}{\Lambda^2}-1  \right) + \frac{9g_X^2m_X^2|_0}{16\pi^2}\left( \ln\frac{m_X^2|_0}{\Lambda^2}+\frac{1}{3}\right) .
    \label{eq:effphimass}
\end{equation}

For the quintessence to explain the dark energy in the late universe, the following conditions should be satisfied at present: 
\begin{equation}
V_{\text{eff}} \sim10^{-123} M_{Pl}^4 \sim 3 \times 10^{-47}  \,\text{GeV}^4 \qquad \text{and} \qquad m_{\phi}\lesssim H_0 \sim 10^{-42} \,\text{GeV} \, ,
\label{eq:conditions}
\end{equation}
where $H_0$ is the Hubble parameter in the present universe.
The first condition comes from the present dark energy density \cite{ParticleDataGroup:2020ssz}. The slow-roll of quintessence requires the second condition. (See App.~\ref{sec:Slow roll condition} for details.)

We demand that each term in Eq.~\eqref{eq:effphimass} is at most the order of $H_0^2$. In principle, each term can be much larger than $H_0^2$, if they cancel each other so that $m_{\phi}^2$ is of the order of $H_0^2$. However, such cancellation can only be possible by fine-tuning, and there is no guarantee that the evolution of $\phi$ during the Hubble time does not break it.

Thus the fourth term in Eq.~\eqref{eq:QV} is much smaller than the third term since $H_0 \ll \Lambda$, and we can ignore the fourth term. The last terms in both Eq.~\eqref{eq:QV} and Eq.~\eqref{eq:effphimass} contain the tree-level dark gauge boson mass $m_X^{}|_0$, so one can constrain the $m_X^{}|_0$ by using the conditions in Eq.~\eqref{eq:conditions}. If we ignore the contribution from the log term, we get $m_X^{}|_0 \lesssim 10^{-11}$ GeV from the last term in Eq.~\eqref{eq:QV} being $\mathcal{O}(10^{-123}M_{Pl}^4)$ or less.
Also, if $m_X^{}|_0 <\Lambda $,  the first derivative of the last term is negative. Hence the quantum correction tends to push the $\phi$ to a larger value. We will discuss further dynamics of the $\phi$ in the next section.

The $m_X^2$ up to the leading order correction is
\begin{equation} 
       m_X^2 = g_X^2 \Big(\phi^2 + \frac{V_0^{\prime\prime}}{32\pi^2}\ln \frac{V_0^{\prime\prime}}{\Lambda^2} \Big) \, .
\end{equation}
The quantum correction to the dark gauge boson mass is negligible since the $V_0^{\prime\prime}$ is of the order of the $H_0^2$ in the present universe.

So far, we have considered $V_0$ as a classical potential. There is an alternative approach taking $V_0$ as the one that already includes all the quantum corrections. However, for the coupled quintessence models, such an approach is less appealing as it requires that any coupling terms of the quintessence should be manipulated to produce the wanted effective potential \cite{Doran:2002bc}. Therefore, it is a fair attitude that one considers the $V_0$ as a classical potential or classical potential + corrections from the quintessence-only loops.
Regardless of which approach is taken, it does not affect the quintessence dynamics in the late time universe for the Ratra-Peebles potential \cite{Doran:2002bc,Brax_2000}, and we take $V_0$ as a classical potential throughout this paper.

\section{Quintessence dynamics}\label{sec:Quintessence dynamics}
The dynamics of the $\phi$ and $X^{\mu}$ are determined from the coupled differential equations in Eq.~\eqref{eq:eom} through the effective potential in Eq.~\eqref{eq:QV}. The equations of motions are connected via 
cosmologically evolving $V_{\text{gauge}}$.

Let us first discuss the sole quintessence dynamics when $g_X=0$, and discuss the $g_X\neq 0$ case later.
When $g_X=0$ (thus $V_\text{gauge} = 0$), the dynamics of the $\phi$ is solely determined by the $\alpha$ and $M$ in Eq.~\eqref{eq:QV}. One can consider the $M$ as a function of $\alpha$, since we adjust the current density of the quintessence to give the right dark energy density \cite{Martin:2008qp}. Therefore, the $\alpha$ and the initial condition are the only parameters determining the excursion of the $\phi$. The potential becomes steeper as the $\alpha$ increases, so the equation of state of the quintessence ($w_0$) at present increases.

\begin{figure}[b]
\begin{subfigure}{0.5\textwidth}
    \includegraphics[width=0.99\linewidth]{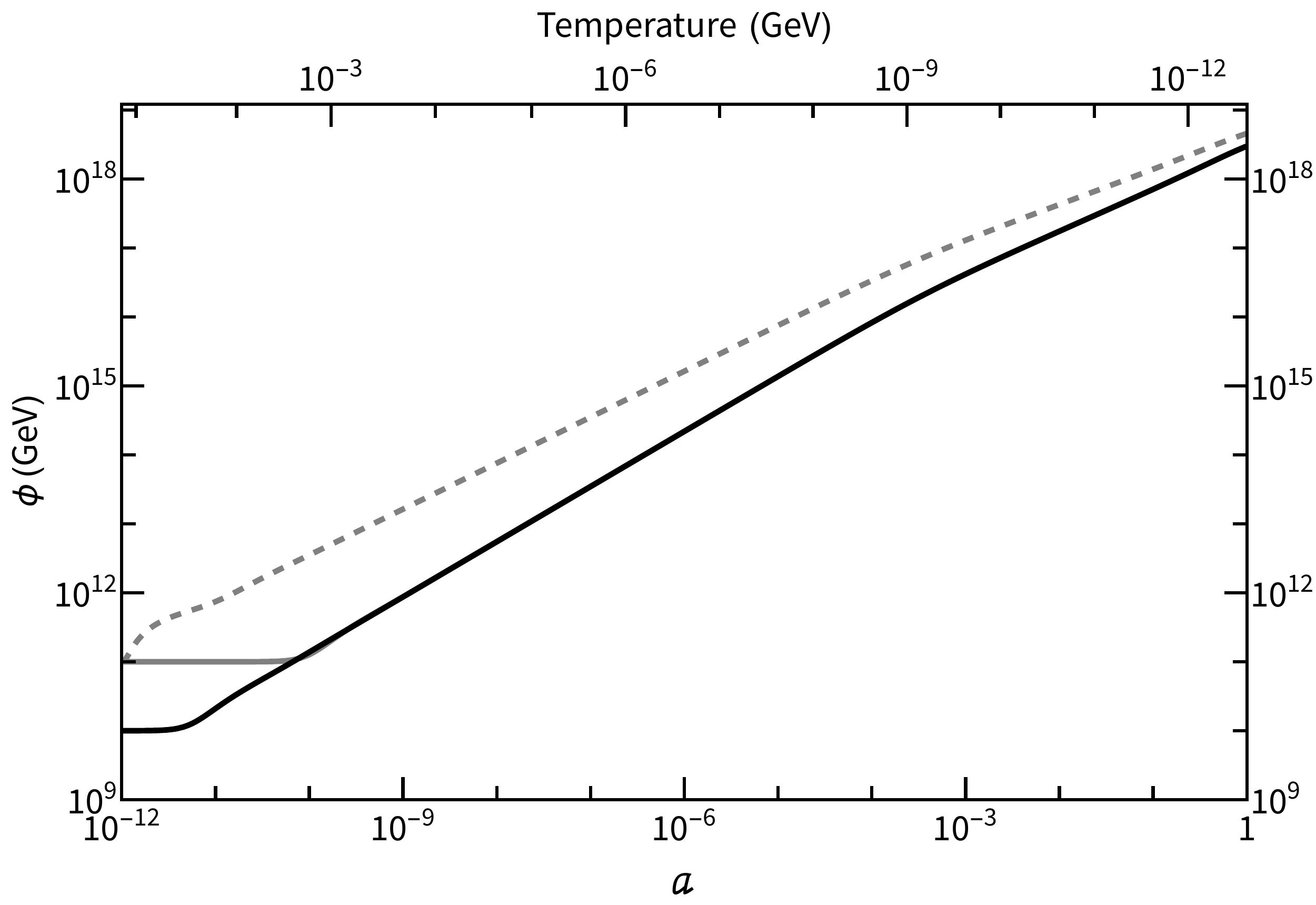}
    \caption{}
\end{subfigure}
\begin{subfigure}{0.5\textwidth}
    \includegraphics[width=0.99\linewidth]{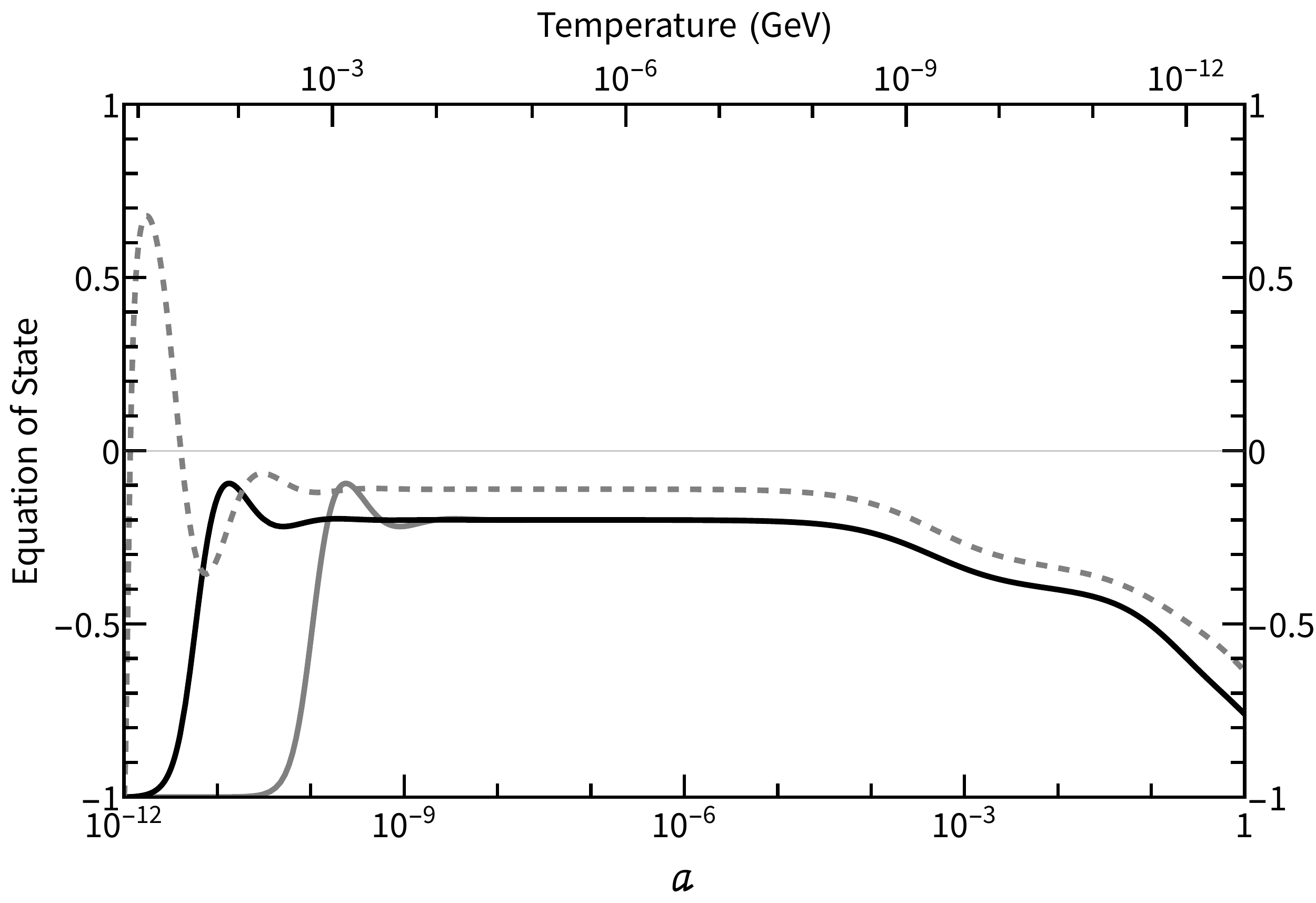}
    \caption{}
\end{subfigure}
    \caption{(a) The $\phi$ excursion for various parameters with (black solid) $\alpha=1$, $\phi_i=10^{10}$ GeV, (gray solid) $\alpha=1$, $\phi_i=10^{11}$ GeV, (gray dashed) $\alpha=2$, $\phi_i=10^{11}$ GeV. (b) Equation of state of quintessence with the same parameter sets. The $\phi_i$ is the initial $\phi$ at $a=10^{-12}$ and $\dot{\phi}=0$. The $M$ is set to give the present dark energy density for $H_0=67.4$km/s/Mpc. It is $M=2.2\times10^{-6}$ GeV for $\alpha=1$, and $M=0.027$ GeV for $\alpha=2$. These plots illustrate the dynamics of the pure quintessence model without the dark gauge symmetry.}
    \label{fig:field_alone}
\end{figure}

It is well-known that the solution of the equation of motion under the Ratra-Peebles potential has a tracking behavior \cite{Steinhardt:1999nw}. Due to the balancing between the slope of potential and the Hubble friction, the solution tries to follow a standard tracking solution. Thus a wide range of the initial condition converges into common late-time behavior. Figure~\ref{fig:field_alone} illustrates that the $\phi$ near the present time is only distinguished by the shape of the potential, i.e., $\alpha$. If the chosen initial value is much smaller or bigger than the tracking value, the $\phi$ can be frozen before it joins the tracking solution. In the former case, the $\phi$ is initially placed on a very steep hill, and it rolls exceedingly quickly so that it overshoots the tracking solution and stays frozen by the Hubble friction until it joins the tracking solution. In the latter case, the potential hill is too shallow to overcome the Hubble friction. In both cases, the $\phi$ joins the tracking solution when $m_{\phi}\sim H$. The solid black and gray curves in Fig.~\ref{fig:field_alone} show that two quintessence fields of different initial field values converge to the tracking solution for $\alpha=1$. On the other hand, the dashed gray curve deviates from the other two by following its tracking solution for $\alpha=2$. One can also see that the equation of state is dynamically varying during the cosmic evolution, and it is closer to $-1$ as the $\alpha$ decreases.

If the $\phi$ follows the tracking solution, $\phi \propto a^{3(1+w_B)/(2+n)}$ when the dominant term in the potential is proportional to $\phi^{-n}$ and $w_B$ is the equation of state of the background \cite{Martin:2008qp,Weinberg:2008zzc}.\footnote{$w_B=1/3$ in the radiation dominated epoch (RD), and $w_B=0$ in the matter dominated epoch (MD).}
In our example, the third term in Eq. (\ref{eq:QV}) dominates when $\phi \ll \sqrt{\alpha(\alpha+1)\Lambda^2/32\pi^2} \sim 10^{18}$ GeV. Hence  the scaling of the $\phi$ is mostly governed by $V_{\rm eff}\propto \phi^{-\alpha-2}$ instead of $\phi^{-\alpha}$.

Now, we discuss the $g_X \neq 0$ case. As we argued in Sec.~\ref{sec:quantum correction}, the dark gauge boson mass should be smaller than $10^{-11}$ GeV. Otherwise, the quantum correction that is proportional to the $m_X^4$ would overproduce the current dark energy density.
Such a light dark gauge boson can be produced in various mechanisms such as coupling to the scalar field \cite{Nakayama:2019rhg,Agrawal:2018vin,Co:2018lka,Bastero-Gil:2018uel,Dror:2018pdh} or gravitational production \cite{Graham:2015rva,Ema:2019yrd,Alonso-Alvarez:2019ixv,Arias:2012az}.\footnote{Note that if such a gauge boson minimally couples to gravity, only the longitudinal mode can be produced through the gravitational effect \cite{Ahmed:2020fhc,Kolb:2020fwh}, whose spectrum is non-thermal and equivalent to the gravitationally produced light scalar \cite{Ling:2021zlj,Garcia:2022vwm,Kaneta:2022gug}. On the other hand, if it couples to the inflaton via the kinetic function of the gauge boson, both longitudinal and transverse modes can be produced during inflation and form a homogeneous condensate \cite{Nakayama:2019rhg}.}

In the rest of this section, instead of specifying a definite mechanism to produce the dark gauge boson, we consider two typical scenarios as effective descriptions of the dark gauge boson at later times. We will show that $V_{\text{gauge}}$ is proportional to the dark gauge boson density in both cases but largely suppressed for the thermal dark gauge boson case. We present a concrete example with the coherent dark gauge boson in Fig.~\ref{fig:gaugedplots}.

We will first consider the dark gauge boson whose distribution is assumed to resemble the thermal one.
Such distribution is possible if it is produced thermally (from either the SM or dark sector heat bath).
In this case, the dark gauge boson in the given mass range is likely to behave as dark radiation due to its small mass.
The second scenario is the dark gauge boson as a coherent state, where the oscillation of the coherent state can be non-relativistic even if its mass is as small as $10^{-30}$ GeV \cite{Nakayama:2019rhg}. This is because the coherent dark gauge boson can be non-relativistic whenever $H \lesssim m_X^{}$. (The $H$ is about $10^{-26}$ GeV around the primordial nucleosynthesis and gradually decreases to the present value of $10^{-42}$ GeV.)
Such a gauge boson can be produced during inflation as a condensate if it couples to the inflaton via the kinetic function of the dark gauge boson \cite{Nakayama:2019rhg,Nakayama:2020rka}.

It should be noted that in both cases we will only consider the case the adiabatic condition \cite{Kofman:1997yn}, which can be written as
\begin{equation}
    \frac{d m_X^{}}{dt} \ll m_X^2 \, ,
    \label{eq:adiabaticity}
\end{equation}
is always satisfied.
If this condition is violated, the WKB-like solution for the wave function of $X_\mu$ cannot be used. In other words, non-perturbative $X_\mu$ production could be non-negligible in such a case.
The adiabatic condition may break in some cases\footnote{For instance, the background quintessence field may roll faster or oscillate abruptly.}, which can bring intriguing phenomenology. 
Furthermore, the violation of the adiabaticity indicates that the fragmentation of a condensate ($\phi$ and/or $X_\mu$) may take place through gauge or self interactions\footnote{The dark gauge boson does not have a tree-level self interaction. Hence, such fragmentation can be made only via higher order corrections.} in a similar manner of other coherent states, such as inflaton \cite{Lozanov:2017hjm} and axion-like particle \cite{Fonseca:2019ypl}.
However, in this paper, we will investigate the simple cases in which this condition is valid.

\subsection{Thermal dark gauge boson} 
The $X_{\mu}X^{\mu}$ in $V_{\text{gauge}}$ can be computed using the Gibbs average \cite{Linde_1979}.
\begin{equation}
\langle X_{\mu}X^{\mu}\rangle= \int \frac{d^3\vec p}{(2\pi)^3} \frac{3 f(\vec p, a)}{E_{\vec p}}  \label{eq:<XX>} \, ,
\end{equation}
where $f(\vec{p},a)$ corresponds to the phase space distribution function including the scale factor. 
Eq.~\eqref{eq:<XX>} is valid under the adiabatic condition given in Eq.~\eqref{eq:adiabaticity}. If the dark gauge boson is produced thermally, it has an effective temperature ($T_f$), which is redshifted from the decoupling temperature ($T_d$) as $T_f=T_d a_d/a$. Then $V_{\text{gauge}} \sim \rho^{}_X m^{2}_X/T_f^2$, and $V_{\text{gauge}}$ is suppressed by $m_X^2/T_f^2$. (See App.~\ref{subsec:Thermally decoupled}.)\footnote{If $m_X > T$, then $V_{\text{gauge}}\sim \rho_{X}$.}

Since the dark gauge boson of our model is extremely light, we expect $V_{\text{gauge}}$ to be largely suppressed. As an example, let us take $g_X=10^{-30}$, which is the largest possible value of the $g_X$ for the tracked quintessence model with the Ratra-Peebles potential (see Fig.~\ref{fig:constraint}), and $\phi \sim 10^{16}$ GeV at $a=10^{-3}$ from the black solid curve in Fig.~\ref{fig:field_alone}. If one assumes that the dark gauge boson once has a similar temperature as the SM heat bath, i.e., $T_f \sim10^{-10}$ GeV, the suppression factor is about $10^{-8}$. Also, the $\rho^{}_X$ should be smaller than the CDM density during the matter dominated era. Hence, $\rho^{}_{X}$ should be smaller than $\rho^{}_{\text{CDM}}$ at the matter-radiation equality, and further diminishes as $\rho^{}_{X} \propto 1/a^4$. 

Due to these reasons, $V_{\text{gauge}}$ cannot affect the $\phi$ dynamics for most of the parameter spaces. However, there is a loophole to overcome the given restrictions. When the dark gauge boson is produced from the dark sector heat bath, its temperature can be much smaller than that of the SM bath (and $m_X$ at the decoupling of the dark gauge boson). Thus the $V_{\text{gauge}}$ can be large enough to affect the quintessence dynamics. Such a possibility could be potentially interesting, but we will not pursue this direction in this paper.

\subsection{Coherent dark gauge boson}
The coherent dark gauge boson \cite{Nakayama:2019rhg} is a spatially homogeneous oscillating field. Therefore, it can be easily described by its equation of motion \cite{Nelson:2011sf}
\begin{equation}
     \ddot{\vec{X}}+H \dot{\vec{X}}+g_X^2\phi^2\vec{X} = 0 \, .
\end{equation}
The dark gauge boson field can have any value prior to inflation. Then the dark gauge boson field takes some random value in the causally connected patch of the universe \cite{Nelson:2011sf}. In the early universe, $H \gg m_X^{}$, so the field is frozen until the oscillation begins when $H\sim m_X^{}$. As the universe cools down, the $H$ can drop below the $m_X^{}$, and the dark gauge boson field is released from the frozen state to the coherently oscillating state. (The detail of the coherent $X$ dynamics is given in App.~\ref{sec:Ultra-light dark gauge boson}.)

\begin{figure}
\begin{subfigure}{0.5\textwidth}
    \centering
    \includegraphics[width=0.99\linewidth]{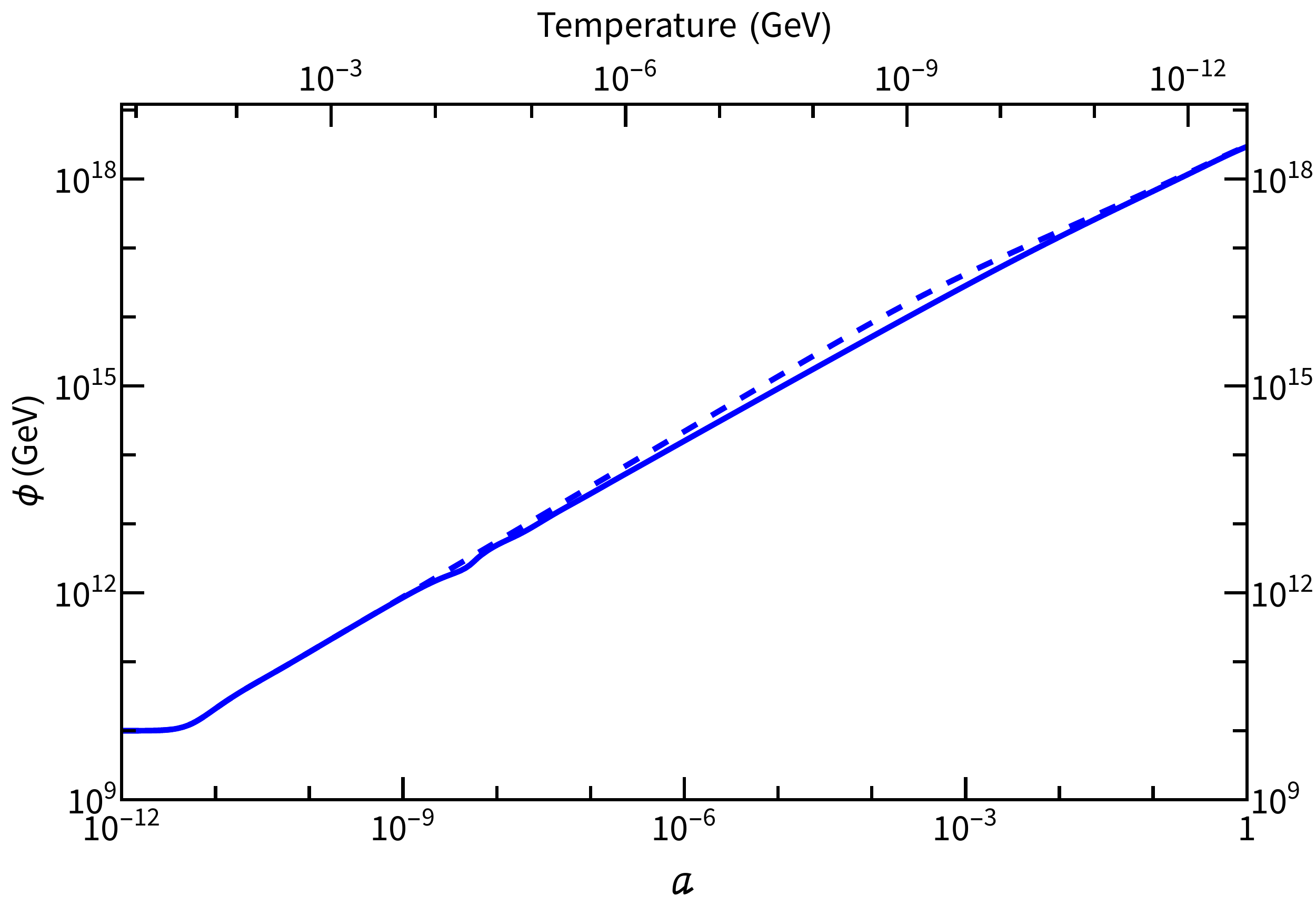}
    \caption{}
    \label{fig:gauge_sub-1}
\end{subfigure}
\begin{subfigure}{0.5\textwidth}
    \centering
    \includegraphics[width=0.99\linewidth]{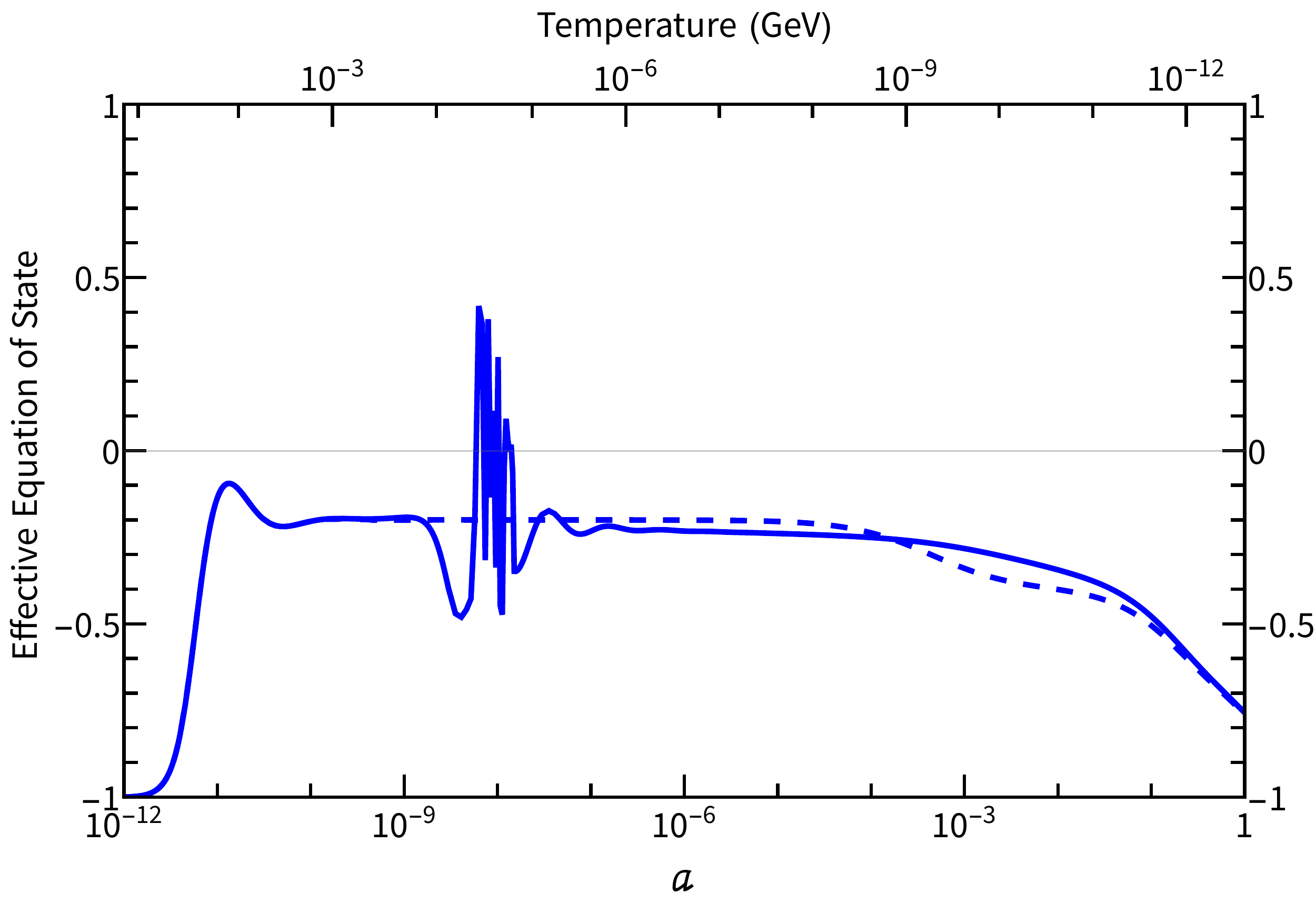}
    \caption{}
    \label{fig:gauge_sub-2}
\end{subfigure}
\newline
\begin{subfigure}{0.5\textwidth}
    \centering
    \includegraphics[width=0.99\linewidth]{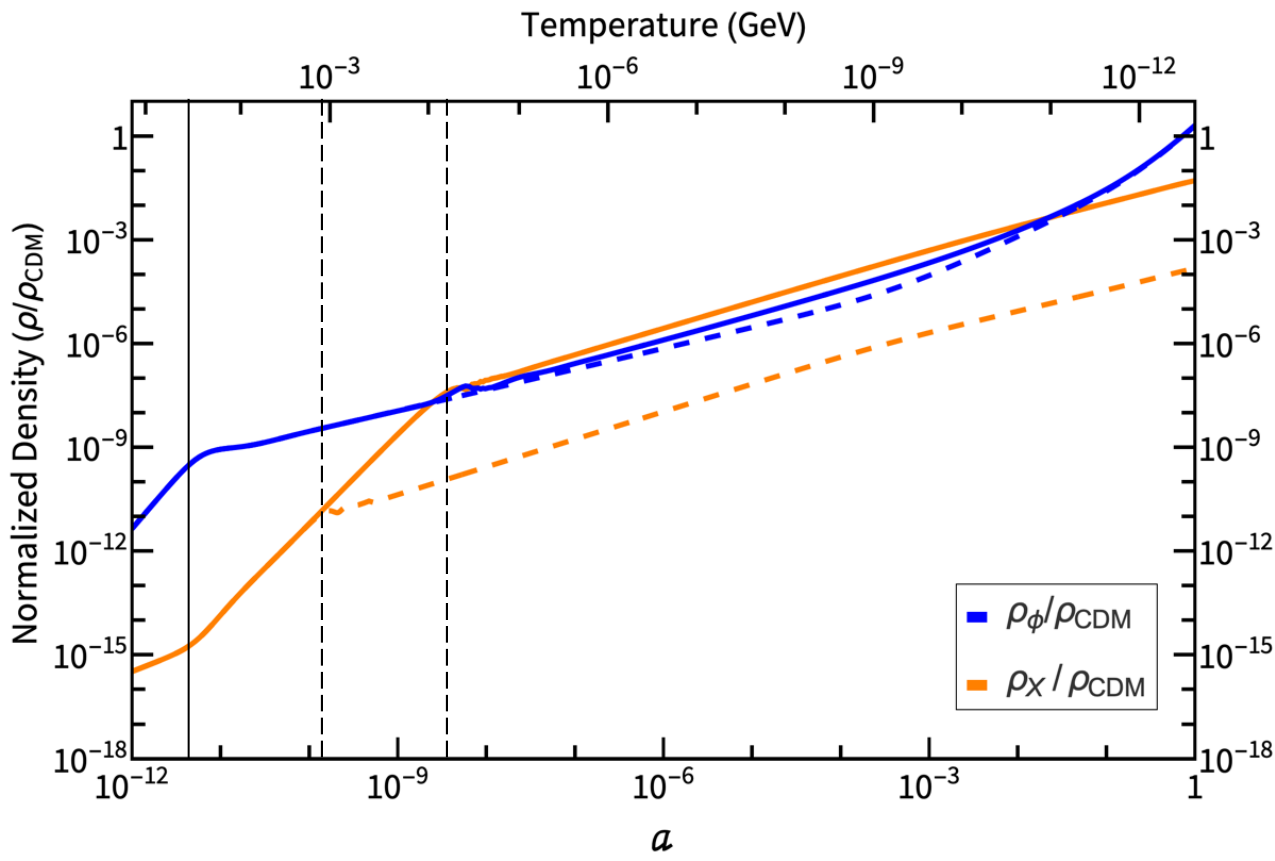}
    \caption{}
    \label{fig:gauge_sub-3}
\end{subfigure}
\begin{subfigure}{0.5\textwidth}
    \centering
    \includegraphics[width=0.99\linewidth]{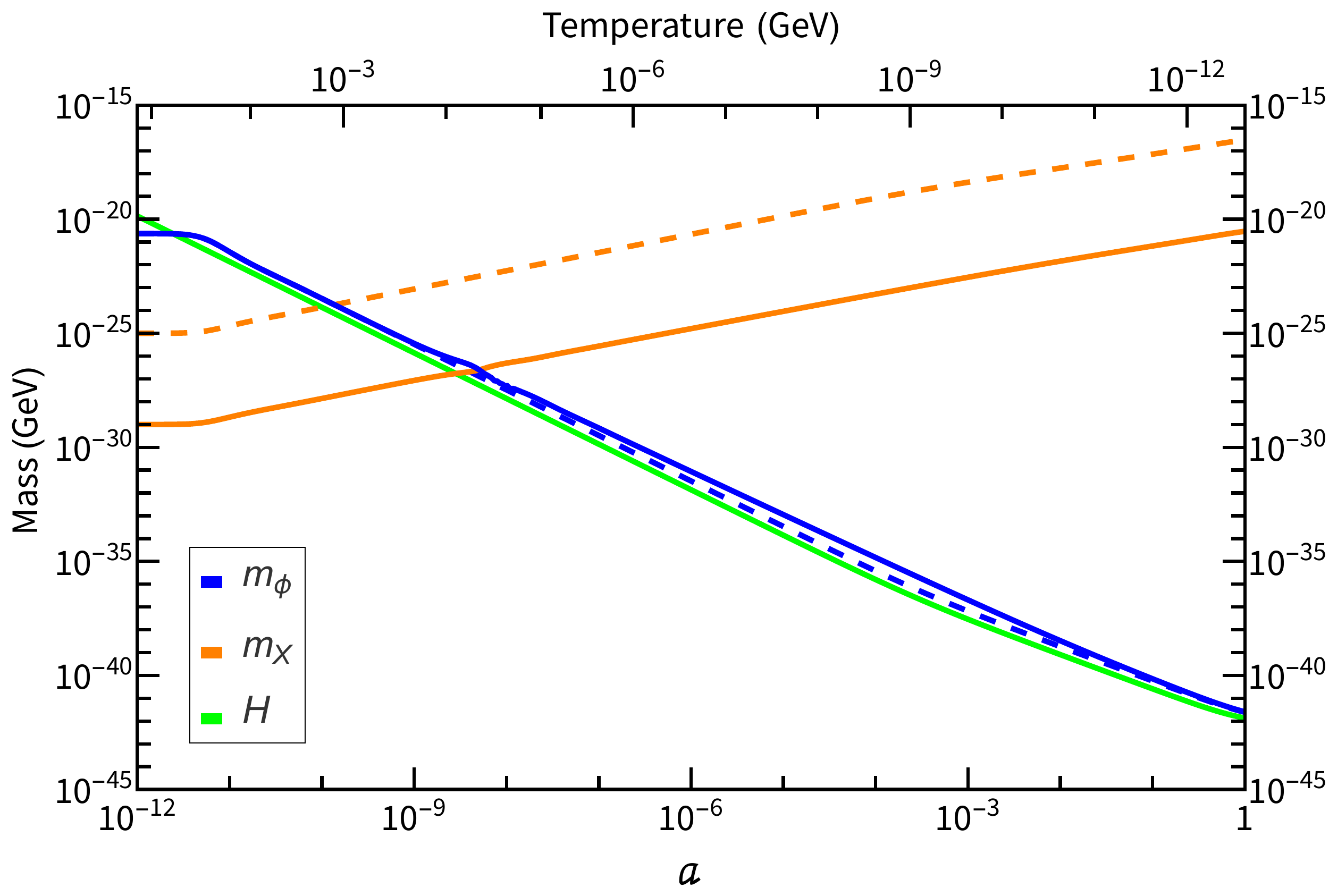}
    \caption{}
    \label{fig:gauge_sub-4}
\end{subfigure}
    \caption{Various plots for the quintessence (blue) with the coherent dark gauge boson (orange) with two choices for the gauge coupling constant: $g^{}_X = 10^{-39}$ (solid), $g^{}_X = 10^{-35}$ (dashed). The plots show (a) the quintessence field values, (b) the effective equation of state of quintessence, (c) the ratio of the dark gauge boson and quintessence energy density to the CDM density, and (d) masses of the dark gauge boson and quintessence. The green curve is the Hubble parameter $H$.
    The shared parameters for all curves at $a=10^{-12}$ are $\alpha=1$, $M=2.2\times 10^{-6}$ GeV, $\Lambda=M^{}_{Pl}$, $\phi_i= 10^{10}$ GeV, $\dot{\phi}_i= 0$, $\rho_X^{}/\rho^{}_{\text{CDM}}|_i= 3 \times 10^{-16}$ (initial fraction of the dark gauge boson energy density to the CDM energy density). $H_0=67.4$ km/s/Mpc and the SM radiation, baryon, and CDM density are taken from Ref.~\cite{ParticleDataGroup:2020ssz}. 
    We assume the CDM is non-relativistic during $a=10^{-12}$ to $a=1$. Note that, in panel (b), the solid curve may not reflect the exact oscillation pattern due to the sampling grid size, but it does not affect any discussions in this paper.} 
    \label{fig:gaugedplots}
\end{figure}

During the oscillation, the dark gauge boson behaves as a non-relativistic matter as it is a condensate of the zero momentum state, and the $X_{\mu}X^{\mu}$ scales as $m_X^{} a^{-3}$. Furthermore, since the $V_\text{gauge} \sim \rho^{}_X$, $V_{\text{gauge}}$ is negligible to the quintessence dynamics if $\rho^{}_X$ is much smaller than the rest of the potential. For instance, when $g_X=10^{-35}$, the $\rho^{}_X$ is always smaller than the quintessence energy density. Hence, $V_{\text{gauge}}$ does not affect the quintessence dynamics, and the $\phi$ excursion and the evolution of the effective equation of state are the same as those shown in the solid black curves in Fig.~\ref{fig:field_alone}. On the other hand, the curves for $g_X=10^{-39}$ show how $V_{\text{gauge}}$ changes quintessence dynamics. As the dark gauge boson energy density becomes comparable to the quintessence energy density, the $\phi$ oscillates about the minimum of the potential until $V_{\text{gauge}}$ becomes subdominant. Such an oscillation is imprinted in the oscillating effective equation of state in Fig.~\ref{fig:gauge_sub-2}. Later on, the dark gauge boson is a subdominant component of the dark matter; hence the quintessence dynamics during the dark energy domination era is not affected by the dark gauge boson. Thus the quintessence recovers the tracking before the dark energy domination era.

\begin{figure}[t]
    \centering
    \includegraphics[width=0.65\linewidth]{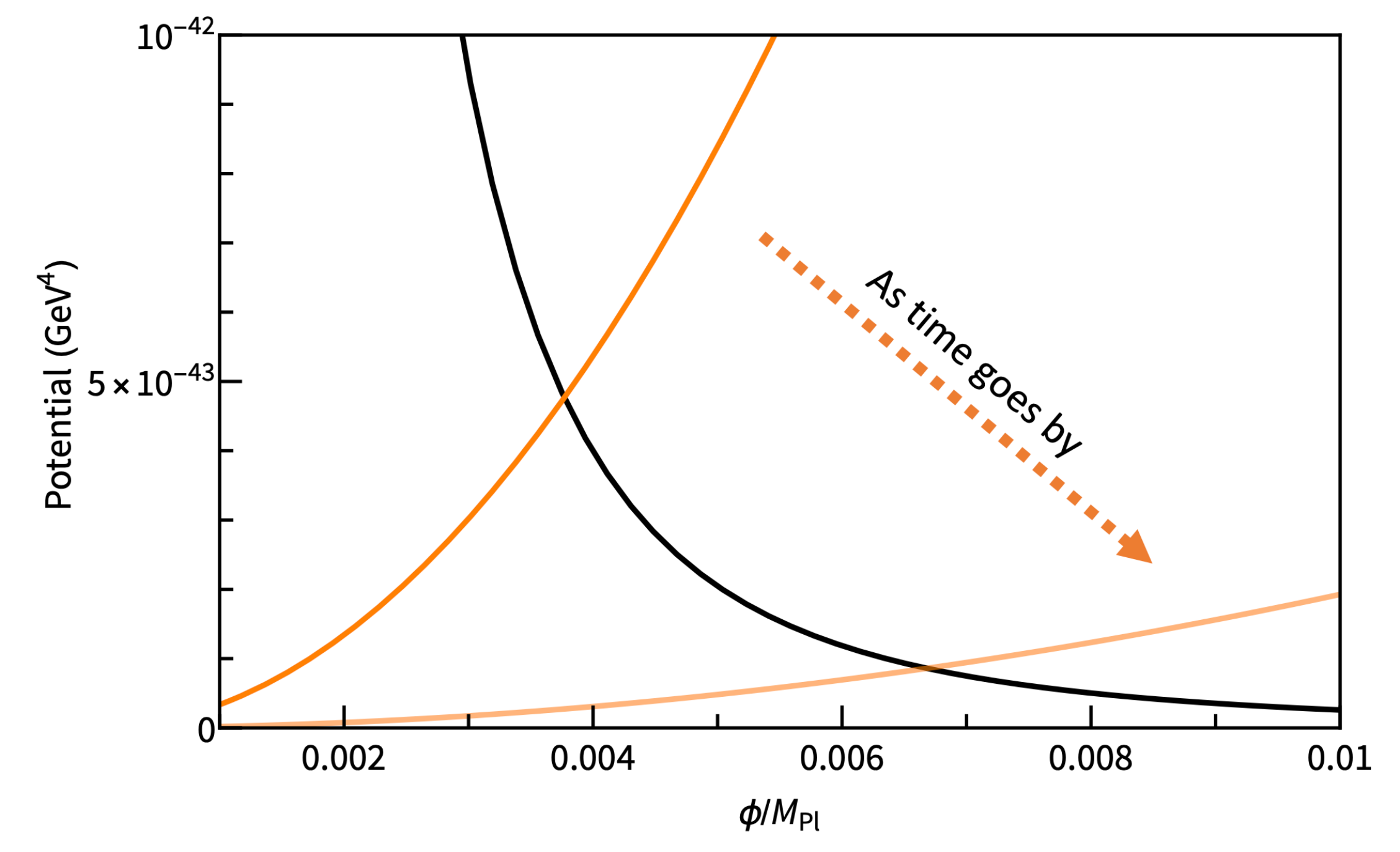}
    \caption{The evolution of the $V_{\text{gauge}}$. The parameter values consistent with the solid curves in Fig.~\ref{fig:gaugedplots} are used. The black curve is the $V_0$ $+$ quantum corrections, and the orange curves are the $V_{\text{gauge}}$. The full effective potential is a sum of the two. The $V_{\text{gauge}}$ evolves from the one at the top (bright orange) to the one at the bottom (light orange) with time.}
    \label{fig:tendency}
\end{figure}

Figure~\ref{fig:tendency} illustrates how $V_{\text{gauge}}$ is significant in some periods but disappears in the late universe. At first, a narrow potential well is made from the combination of $V_{\text{gauge}}$ and other terms in the potential. So the $\phi$ oscillates about the minimum of the potential ($\phi_{\text{min}}$) and does not follow the tracking solution of the sole Ratra-Peebles potential. Then, as $V_{\text{gauge}}$ is redshifted away, the $\phi_{\text{min}}$ also moves toward the larger value, and the potential well becomes shallower. Hence, the oscillation cannot persist, the $\phi$ cannot track the $\phi_{\text{min}}$ anymore, and $\phi$ is left behind. Then $V_{\text{gauge}}$ does not affect the $\phi$ dynamics anymore, and the $\phi$ joins the tracking solution. As an example, the solid curve in Fig.~\ref{fig:gauge_sub-1} shows the deviation from the tracking solution due to $V_{\text{gauge}}$ but accordance in the late universe.

As we mentioned, the dynamics of $\phi$ and $X$ drastically change when the hierarchy between the $H$ and the mass of the each field is switched. This is essential to understand the evolution of $\rho^{}_X$. Figure~\ref{fig:gauge_sub-4} compares $m_X^{}$, $m_{\phi}$, and $H$. The $m_X$ is simply proportional to the $\phi$, but the $m_{\phi}\sim H$ due to the tracking.\footnote{The $m_{\phi}$ can be larger than $H$ if $V_{\text{gauge}}$ hinders the growth of $\phi$.} The moment when the $m_{\phi}$ ($m_{X}$) crosses with the $H$ is marked by the solid (dashed) vertical lines in Fig.~\ref{fig:gauge_sub-3}. One can easily see that those vertical lines coincide with the sudden warp of $\rho_X^{}$.
The followings are brief descriptions of Fig.~\ref{fig:gauge_sub-3}.\footnote{When 
$m_{\phi} < H$, $m_{X} > H$ : $X$ is in the coherent oscillation, but $\phi$ is frozen. So, $\rho_X^{} \propto a^{-3}$. However, this setting does not occur in our example.}
\begin{itemize}
    \item $m_{\phi} < H$, $m_{X} < H$ : Both $X$, $\phi$ are frozen by the Hubble friction. So, $\rho_X^{} \propto a^{-2}$. This regime is the left side of the first bending of orange curves. 
    \item $m_{\phi} > H$, $m_{X} < H$ : $X$ is frozen, but $\phi$ is running on the potential. So, $\rho_X^{} \propto \phi^2 a^{-2} \propto m_X^2 a^{-2}$.  This regime is between the first and second bending of orange curves. ($\rho_X^{} \propto a^{-2/5}$ and $\rho_X^{}/\rho_{\text{CDM}}\propto a^{13/5}$ in the RD, or $\rho_X^{} \propto a^{-4/5}$ and $\rho_X^{}/\rho_{\text{CDM}}\propto a^{11/5}$ in the MD.) 
    \item $m_{\phi} > H$, $m_{X} > H$ : $X$ is in the coherent oscillation, and $\phi$ is running on the potential. So, $\rho_X^{} \propto \phi a^{-3} \propto m_X^{} a^{-3}$. The factor of $a^{-3}$ is interpreted as a dilution of number density by the expansion of the universe, and the additional factor of $m_X^{}$ is an energy of an individual dark gauge boson. We can identify the dark gauge boson as mass varying CDM. This regime is the right side of the second bending 
    of orange curves.  ($\rho_X^{} \propto a^{-11/5}$ and $\rho_X^{}/\rho_{\text{CDM}}\propto a^{4/5}$ in the RD, or $\rho_X^{} \propto a^{-12/5}$ and $\rho_X^{}/\rho_{\text{CDM}}\propto a^{3/5}$ in the MD.)
\end{itemize}

Note that if the conditions other than the gauge coupling are the same, the present dark gauge boson density is proportional to $g_X^2$ as one can see from Eq.~\eqref{eq:condensatedensity}. Nevertheless, the present dark gauge boson energy density in Fig.~\ref{fig:gauge_sub-3} increases as the $g_X$ decreases. This is because the curves are drawn with a fixed initial dark gauge boson density ($\rho_X^{}/\rho_{\text{CDM}}|_i$), not a fixed initial $X$ field value. If one decreases the $g_X$ with the fixed initial $X$ field value, the present energy density of the dark gauge boson also decreases. Likewise, it explains why the oscillation of the $\phi$ gets stronger as $g_X$ decreases, as one can see from Fig.~\ref{fig:gauge_sub-2}. 

One should note that the dark gauge boson may constitute an extremely tiny portion of the total energy density of the universe in the early universe, but it grows to take a significant portion in the present universe. For example, the normalized energy density of the solid orange curve in Fig.~\ref{fig:gauge_sub-3} is $3\times 10^{-16}$ initially but is an order of $0.01 - 0.1$ in the present universe. This is because the $\phi$ can grow by $10^8$ times larger than its initial value, and the energy density of the dark gauge boson before the coherent oscillation begins scales as $a^{-2}$ while the CDM scales as $a^{-3}$. If the energy density of the dark gauge boson is comparable to the quintessence energy density in the recent universe, it can affect the expansion of the universe. In such a case, one has to consider the equation of state of the $\phi+X$ fluid. 

Considering that the energy density of the dark gauge boson evolves as mass varying CDM in the late universe ($p_X=0$), using $a^3 \rho_X^{} / m_X^{} = \rho_X^0 / m_X^0$ in the non-relativistic limit. Often it is convenient to define the effective dark energy density and effective CDM density \cite{Das:2005yj}, as
\begin{equation}
\begin{split}
\rho_{\widetilde{DE}} &\equiv \rho_{\phi} + \left( \frac{m^{}_X}{m_X^{0}}-1\right) \frac{\rho_X^{0}}{a^3} \, , \\ \rho_{\widetilde{CDM}}&\equiv \frac{\rho^0_X+\rho^0_{\text{CDM}}}{a^3} \, ,
\end{split}
\label{eq:effrho}
\end{equation}
with the Friedmann equation
\begin{equation}
3m_{Pl}^2 H^2=\rho^{}_{\widetilde{DE}}+\rho_{\widetilde{CDM}} +\rho_b \, ,
\end{equation}
where we suppose that $\rho^0_X+\rho^0_{\text{CDM}}$ accounts for the total dark matter energy density observed in the present universe. The quantities with the superscript zero denote the present values, and the $\rho_b$ is the baryon density. 
Here, one can separate the $a^{-3}$ scaling and only include mass varying effects to the dark energy component. Therefore, it is easy to compare with the numerical results, which usually assume non-interacting CDM. By taking the time derivative of Eq.~\eqref{eq:effrho} and using Eqs.~\eqref{eq:eom}, \eqref{eq:phidensandP}, and \eqref{eq:vgaugecoherentosc}, the effective equation of state of the effective dark energy density $w_{\text{eff}}(\widetilde{DE})$ can be obtained \cite{Das:2005yj}. (Replace the $\rho_{\phi}$ in Eq.~\eqref{eq:findeos} with the $\rho_{\widetilde{DE}}$.)
\begin{equation}
\label{eq:rhoDE}
\begin{split}
    \dot{\rho}_{\widetilde{DE}}&= \dot{\rho}_{\phi}-3H\left(\frac{m^{}_X}{m^{0}_X}-1\right)\frac{\rho^{0}_X}{a^3}+\frac{\dot{m}^{}_X}{m^0_X}\frac{\rho_X^0}{a^3}\\
    &= -3H\left((1+w_0)\rho_{\phi}+ \left(\frac{m^{}_X}{m_X^0}-1\right)\frac{\rho^0_X}{a^3} \right) \\
    &=-3H \left(1+w_{\text{eff}}(\widetilde{DE})\right)\rho_{\widetilde{DE}} \, .
\end{split}
\end{equation}
Then the effective equation of state for the effective dark energy density is given by
\begin{equation}
   w_{\text{eff}}(\widetilde{DE})= -1 + \frac{1}{\rho_{\widetilde{DE}}}\left( (1+w_0)\rho_{\phi} + \left(\frac{m^{}_X}{m^0_X}-1\right)\frac{\rho_X^{0}}{a^3} \right) .
\label{eq:effwDEeff}
\end{equation}
A condition $p_X=0$ is used in the second equality. A simple calculation shows that $p_{\phi}=w_0\rho_{\phi}=w_{\text{eff}}(\widetilde{DE})\rho_{\widetilde{DE}}=p_{\widetilde{DE}}$, which tells that the pressure of the effective dark energy is solely determined by the pressure of the quintessence since $p_X=0$.

\begin{figure}[t]
    \centering
    \includegraphics[width=0.65\linewidth]{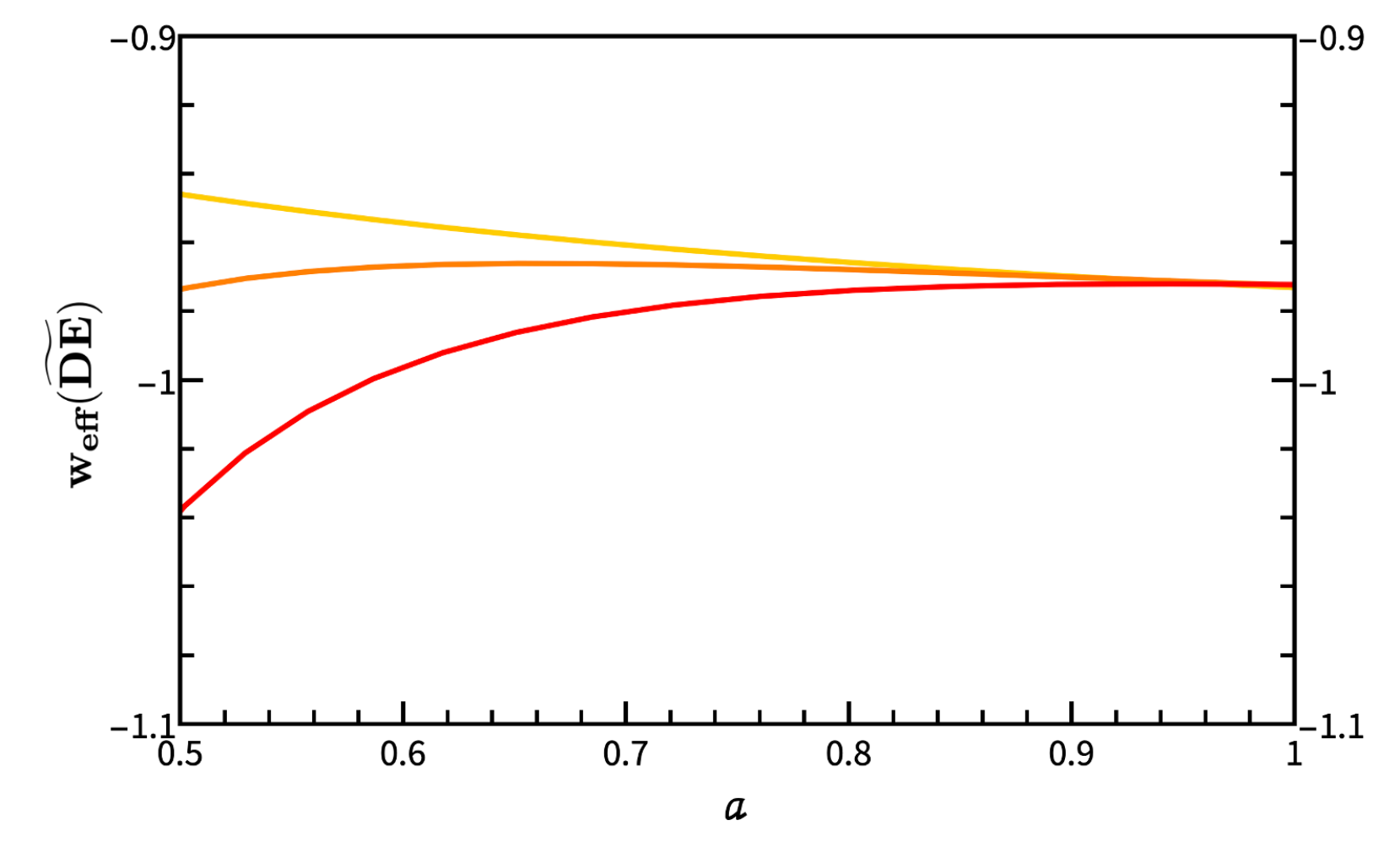}
    \caption{The $w_{\text{eff}}(\widetilde{DE})$ in the gauged quintessence model with various amounts of the dark gauge boson density. The present energy density fractions of the dark gauge boson over the CDM are 0.013 (yellow curve), 0.09 (orange curve), 0.27 (red curve). The uncoupled quintessence model would nearly overlap with the yellow curve. The $\alpha=1/16$, $M=6.3\times10^{-12}$ GeV, $g_X=10^{-39}$ are common for all curves. The energy of dark gauge boson scales as $\propto m_X/a^3$ and evolution of $\phi$ is the same for all cases in the given domain.
    }
    \label{fig:wDEeff}
\end{figure}

\section{Gauged quintessence on the Hubble tension}\label{sec:Gauged quintessence on the $H_0$ tension}
The 5$\sigma$ discrepancy between the \emph{Planck} satellite result \cite{Planck2020} and the distance ladder measurement \cite{Riess:2021jrx} on the Hubble constant is referred to as the Hubble tension. It is argued that the uncoupled quintessence model may worsen the Hubble tension \cite{Banerjee:2020xcn,Lee:2022cyh}.

In order to alleviate the Hubble tension, we need $w_{\text{eff}}(\widetilde{DE})<-1$ \cite{Heisenberg:2022lob,Heisenberg:2022gqk,Lee:2022cyh,DiValentino:2017iww,DiValentino:2016hlg,Joudaki:2017zhq}.
This is because any late universe change in the cosmic model should preserve the comoving distance since the baryon acoustic oscillation angular scale ($\theta_s$) is a model-independent quantity \cite{Heisenberg:2022lob,Heisenberg:2022gqk}. In other words, if the sound horizon ($r_{s}$) does not change, then the angular diameter distance ($D_A$) to the last scattering should be fixed.
\begin{equation}
    D_A = \frac{r_s}{\theta_s} = \int^1_{a_s}  \frac{da}{a^2H(a)} \, ,
\end{equation}
where the $a_s$ is the scale factor at the last scattering. Hence the larger $H_0$, which can alleviate the Hubble tension, should be compensated by the smaller $H$ in the recent past. Since $H \approx \sqrt{\rho_{\widetilde{DE}}/(3m_{Pl})^2}$ in dark energy dominated era, $\rho_{\widetilde{DE}}$ should increase over time to relieve the Hubble tension. This condition demands $w_{\text{eff}}(\widetilde{DE})<-1$ for some $z$ \cite{Heisenberg:2022lob,Heisenberg:2022gqk}.

In this sense, the gauged quintessence model can perform better than the uncoupled quintessence model. If $\dot{m}_X>0$ (i.e. $\dot{\phi}>0$) and there are sufficient dark gauge boson energy density, the $w_{\text{eff}}(\widetilde{DE})$ becomes lower than that of the uncoupled quintessence model. The decrease of the $w_{\text{eff}}(\widetilde{DE})$ becomes larger if (i) the present energy density of the dark gauge boson is more significant, (ii) the change of the dark gauge boson mass is bigger. [See Eq.~\eqref{eq:effwDEeff}.] This can be observed in Fig.~\ref{fig:wDEeff}. In the past, the red curve shows the lowest $w_{\text{eff}}(\widetilde{DE})$ as its present dark gauge boson energy density is the largest among the three cases. All three curves have the same $w_{\text{eff}}(\widetilde{DE})$ in the present since the dark gauge boson contribution vanishes.

\section{Constraint on the dark gauge coupling}\label{sec:cons_on_gauge}
\begin{figure}[b]
    \centering
    \includegraphics[width=0.65\linewidth]{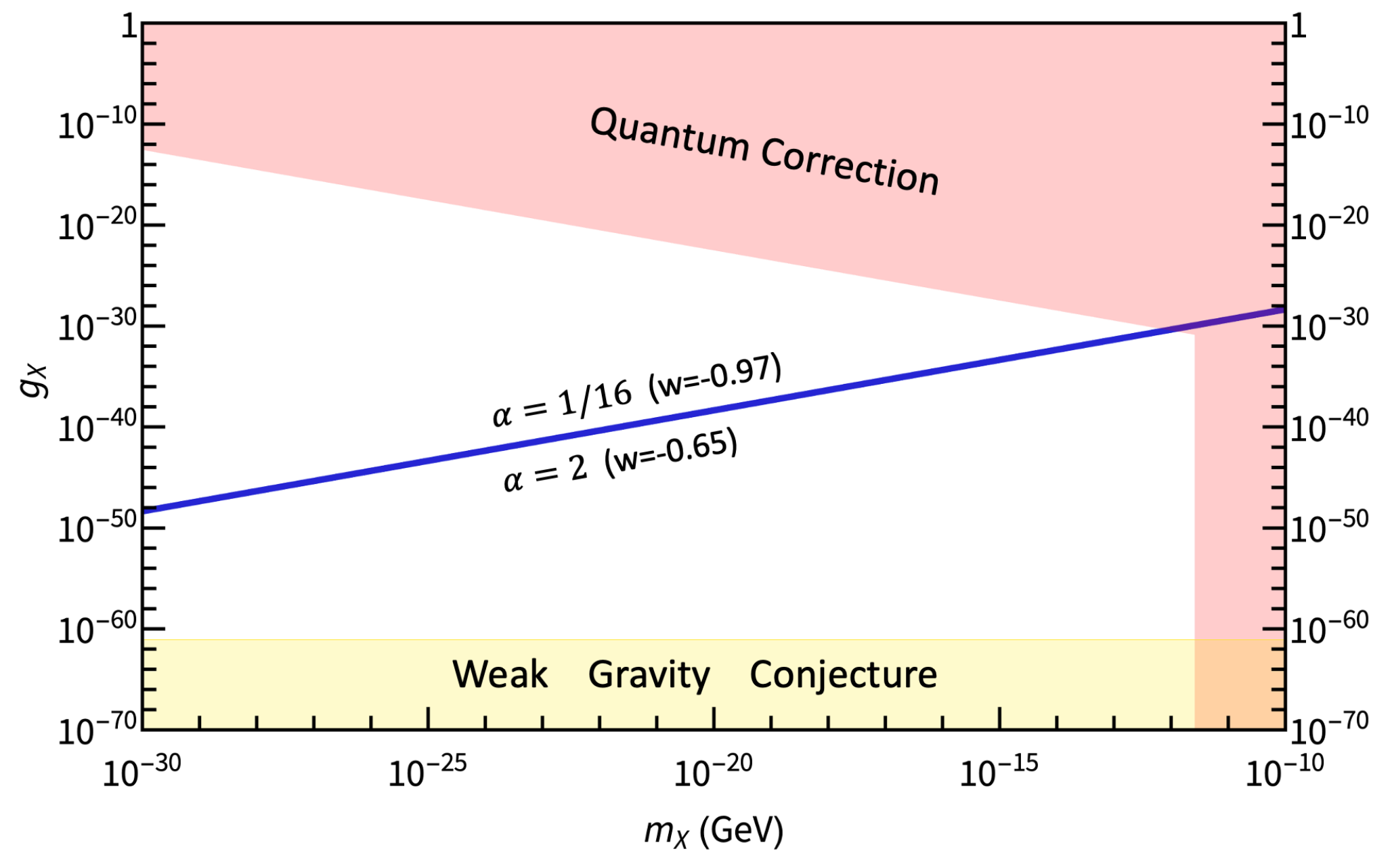}
    \caption{The solution and constraints on the dark gauge coupling $g_X$ with a present dark gauge boson mass $m_X^{}$. The blue band is the allowed region for the model with the Ratra-Peebles potential with $\alpha=1/16$ to $\alpha=2$. This $\alpha$ range is where the quintessence shows the tracking behavior. The conditions in Eq.~\eqref{eq:conditions2} exclude the red area. The yellow area is disfavored by the weak gravity conjecture \cite{Arkani-Hamed:2006emk}. The cutoff $\Lambda = M_{Pl}$ is adopted.} 
    \label{fig:constraint}
\end{figure}

As discussed in Sec.~\ref{sec:quantum correction}, the quantum correction from the gauge boson loops may ruin the dark energy behavior of the quintessence. Order of magnitude estimation of the constraints on the parameter space can be obtained if one demands the criteria that the magnitude of the gauge boson corrections for $V_{\text{eff}}$ and $m_{\phi}$ [the last terms in Eqs.~\eqref{eq:QV} and \eqref{eq:effphimass}] at present are smaller than the other terms.
\begin{equation}
\begin{split}
&\text{(i}) \quad \frac{3(m_{\text{X}}^2|_0)^2}{64\pi^2} \left(\ln\frac{m_{\text{X}}^2|_0}{\Lambda^2}-\frac{5}{6} \right)< 3 \times 10^{-47}  \,\text{GeV}^4 \, , \\
&\text{(ii}) \quad \frac{9g_X^2m_X^2|_0}{16\pi^2}\left( \ln\frac{m_X^2|_0}{\Lambda^2}+\frac{1}{3}\right) < 10^{-84} \,\text{GeV}^2 \, .
\end{split}
\label{eq:conditions2}
\end{equation}
The criteria $\text{(i)}$ only depends on the $m_X^{}$. Thus it constrains the $m_X^{}$. (Note $m^2_X \approx m^2_X|_0$ in the present universe as discussed in Sec.~\ref{sec:quantum correction}.)
The criteria $\text{(ii)}$ depends on both $m_X^{}$ and $g_X$, and it constrains the $g_X$ at each $m_X^{}$. In Fig.~\ref{fig:constraint}, criteria (i) constrains the right red region, and criteria (ii) constrains the upper red region. These constraints only account for the quantum correction, and whether dark gauge bosons are particles or coherent is irrelevant.

The weak gravity conjecture \cite{Arkani-Hamed:2006emk} states that gravity is the weakest force in nature, which requires
\begin{equation}
    \frac{m}{M_{Pl}} \lesssim g_X \, ,
\end{equation}
where $m$ is the mass of the lightest charged particle. In our case, since the quintessence is the only one charged by the $U(1)_\text{Dark}$ gauge symmetry, it is the quintessence mass $m_{\phi}$. Thus it constrains the $g_X$.  

The $g_X$ and $m_X^{}$ are not independent as $m_X^{} \approx g_X \phi$, and the present day $\phi$ value is determined by the excursion of $\phi$ that depends on the shape of the potential and initial conditions. Furthermore, the Ratra-Peebles potential has a tracking behavior, so a wide range of the initial conditions converges to the common late-time solution \cite{Steinhardt:1999nw}. This is even true under the presence of the $V_{\text{gauge}}$, if $V_\text{gauge}$ is subdominant in the recent universe. Therefore, the actually allowed parameter space of the model with a specific potential may be limited.

The blue region in Fig.~\ref{fig:constraint} shows the parameter space with the Ratra-Peebles potential that shows the tracking behavior. 
While there is wider parameter space if we do not require the tracking behavior, those with the tracking would not stretch much from the given blue region (with $w_\text{eff} = -0.97$ to $-0.65$) when $\alpha$ is an order of $0.1 - 1$.
There is a one-to-one correspondence between the $\alpha$ and the present equation of state since the present equation of state of the quintessence with the Ratra-Peebles potential only depends on $\alpha$ if the gauge boson induced terms are small compared to the $V_0$, and the quintessence is under the tracking.
It is known that the $\phi\sim M_{Pl}$ if the quintessence field is driven by the Ratra-Peebles potential. Thus, although we showed only the band bounded by two representing choices of the $\alpha = 1/16$ ($w_\text{eff} = -0.97$) and $2$ ($w_\text{eff} = -0.65$), the curves with other choices would lie close to the blue band. The blank area in Fig.~\ref{fig:constraint} shows the viable parameter space independent of shapes of the potential.
 
The quantum correction constraints in Eq.~\eqref{eq:conditions2} originate from the quartic interactions between the quintessence and the dark gauge boson. Thus, the constraint from the quantum correction is independent of the detail of the quintessence model (i.e., the form of $V_0$). In this sense, the red constraints of Fig.~\ref{fig:constraint} are model-independent.

We evaluated the weak gravity conjecture with the present quintessence mass. However, the constraint may differ if one takes the quintessence mass at different times, which will be larger than the present value. Therefore, the weak gravity conjecture bound in Fig.~\ref{fig:constraint} has to be interpreted as a conservative bound of the gauge coupling in the presence of the weak gravity conjecture.

\section{Summary and outlook}\label{sec:Summary}
In this paper, we investigated to answer the question ``What happens if the dark energy component of the universe is under a gauge symmetry?''.
Dark energy is garnering more and more attention these days; it is the most elusive part of the universe, and the observation data regarding dark energy are becoming precise enough to test the dark energy models.
We assume the dark energy is dynamic rather than a constant de Sitter and take the popular Ratra-Peebles quintessence scalar model as a basic form.
We generalized it to a complex scalar to impose an Abelian gauge symmetry and presented the gauged quintessence scalar as a new cosmological dark energy model.

We studied the general properties of the model in both qualitative and quantitative ways.
We found that the gauge symmetry in the dark energy sector brings intriguing characters to the dark energy model in a very general way.
First, the dark energy complex scalar field provides mass to the dark gauge boson.
Second, the dark gauge boson mass and quintessence scalar mass are cosmically evolving.
Third, the new gauge symmetry provides extra contributions to the quintessence mass and the scalar potential by quantum corrections due to the new gauge boson.
This imposes severe constraints on the new gauge symmetry for the dark energy field since both the quintessence mass and the scalar potential at the present universe should satisfy certain conditions for a viable dark energy candidate.
Using these constraints, we obtained the upper bound on the new gauge coupling constant and found it is very small, but it can still be consistent with the weak gravity conjecture.
The constraint on the gauge coupling constant limits the dark gauge boson mass to be tiny ($m_X^{} \lesssim 10^{-12}$ GeV).
Moreover, our study suggests that applying the weak gravity conjecture, which involves both the coupling and mass of a particle may not be naive when the mass is varying cosmologically.
We also discussed the oscillating feature of the quintessence scalar due to the gauge symmetry and how it fades away with time in our scenario.
It is also interesting to note that the dark gauge boson contribution to the effective equation of state of the dark energy may alleviate the Hubble tension issue suffered greatly by the original quintessence model.

Although we studied only a simple scenario where there is only a gauge boson with a small coupling constant, there are potential directions to extend the scenario.
For instance, the gauge symmetry may be extended to the dark matter sector connecting the dark energy and the dark matter with a dedicated interaction in the dark sector.
For this to be valid, however, the flatness of the scalar potential should be arranged by some symmetries or mechanisms.
The kinetic mixing between the dark gauge boson and the photon, which is not forbidden by any symmetry, can be another potentially interesting direction.
Considering dark matter searches in both direct and indirect ways are very active partly because of their presumed non-gravitational interactions, the introduction of a new interaction in the dark energy sector may also enrich the dark energy studies.

\acknowledgments
We thank H. Park and M. Seo for the discussions. We thank E.O. Colgain for useful comments on the relationship between the quintessence model and the Hubble constant.
This work was supported in part by the JSPS KAKENHI (Grant No. 19H01899) and the National Research Foundation of Korea (Grant No. NRF-2021R1A2C2009718).

\appendix

\section{Slow-roll condition}\label{sec:Slow roll condition}
The slow rolling of the quintessence scalar is essential in typical quintessence models for dark energy. In this appendix, we briefly review how the quintessence mass is related to the slow-roll condition in Eq.~\eqref{eq:conditions}. We do not restrict the potential to be Ratra-Peebles potential here. Instead, the discussion applies to any potential, which gives the slow rolling in the late universe.

The equation of motion of the $\phi$ with an arbitrary potential can be written as
\begin{equation}
    \ddot{\phi}+3 H\dot{\phi} + \frac{\partial V_{\text{eff}}(\phi)}{\partial \phi} =0 \, ,
    \label{eq:apendxeom}
\end{equation}
where the $V_{\text{eff}}(\phi)$ is the effective potential which includes the thermal and quantum corrections. For the slow rolling to be stable during the Hubble time, $\Delta t \sim 1/H$, which is the typical time scale of the dark energy domination, it should be satisfied that
\begin{equation}
   |\ddot{\phi}| < |H \dot{\phi}|\, .
   \label{eq:appendineq1}
\end{equation}

Hence, Eq.~\eqref{eq:apendxeom} is reduced to
\begin{equation}
   3 H\dot{\phi} + \frac{\partial V_{\text{eff}}(\phi)}{\partial \phi} \approx 0 \, ,
\end{equation}
Further, one can take the time derivative on the above equation and arrange the terms as
\begin{equation}
    \frac{\partial^2 V_{\text{eff}}(\phi)}{\partial \phi^2}  \approx -3 H\frac{\ddot{\phi}}{\dot{\phi}}-3 \dot{H}
    \lesssim 3 (H^2+|\dot{H}|) \, ,
\label{eq:appendmiddle}
\end{equation}
using Eq.~\eqref{eq:appendineq1}. The last step is to use the following Friedmann equations
\begin{equation}
\begin{split}
    H^2&= \frac{1}{3m_{Pl}^2}\left( \frac{1}{2}\dot{\phi}^2+V_{\text{eff}}(\phi)+\rho_m+\rho_r\right) \, , \\ |\dot{H}|&=\left| \frac{1}{2m_{Pl}^2}\Big(\dot{\phi}^2 +\rho_m+\frac{4}{3}\rho_r\Big)\right| \lesssim \left| \frac{1}{2m_{Pl}^2}\Big(\frac{1}{2}\dot{\phi}^2+V_{\text{eff}}(\phi) +\rho_m+\frac{4}{3}\rho_{r}\Big) \right| \sim \frac{3}{2}H^2 \, ,
\end{split}
\label{eq:apendfriedmann}
\end{equation}
where the $\rho_m$ and $\rho_r$ are the matter and radiation density, and the inequality of the second equation holds since the dark energy equation of state is less than $-1/3$ (i.e., $\dot{\phi}^2 < V_{\text{eff}}(\phi)$), and the $\rho_r$ is much smaller than the other components.

Finally, one can combine Eqs.~\eqref{eq:appendmiddle} and \eqref{eq:apendfriedmann} to get
\begin{equation}
     \frac{\partial^2 V_{\text{eff}}(\phi)}{\partial \phi^2} = m_{\phi}^2 \lesssim \frac{15}{2} H^2 \, .
\end{equation}
This inequality states that the quintessence mass should be comparable to or smaller than the Hubble scale to have a slow rolling. In the main text, we simply state that $m_{\phi}\lesssim H$. [See Eq.~\eqref{eq:conditions}].

\section{Evolution of the gauge potential}\label{sec:Redshift of the gauge potential}
In this appendix, we demonstrate the evolution of the $V_\text{gauge}$ by the redshift without any particle to particle interactions.

\subsection{Thermally decoupled}\label{subsec:Thermally decoupled}
The $ \langle X_{\mu}X^{\mu} \rangle $ is given from Eq.~\eqref{eq:<XX>},
\begin{equation}
\langle X_{\mu}X^{\mu} \rangle  =3 \int\frac{d^3\vec p}{(2\pi)^3} \frac{ f(|\vec p|,a)}{\sqrt{m_X^2+|\vec p|^2}} \, ,
   \label{eq:Vgauge_1}
\end{equation}
where the scale factor dependence of the phase space is explicitly written. Due to the expansion of the universe, the momenta of the particles are redshifted. Hence, if one knows the phase space density at the specific scale factor `$a_0$', one could find the phase space density at the arbitrary scale factor `$a$' by the relation $f(p,a_0)=f(a_0 p/a,a)$. 

Therefore, the $ \langle X_{\mu}X^{\mu} \rangle $ and the $V_{\text{gauge}}$ can be written as
\begin{equation}
\langle X_{\mu}X^{\mu} \rangle  =3\left(\frac{a_d}{a}\right)^3 \int\frac{d^3 \vec p}{(2\pi)^3} \frac{ f( |\vec p|,a_d)}{\sqrt{ m_X^2+\left(\frac{a_d}{a}\right)^2|\vec p|^2}} \, ,
  \label{eq:Vgauge_form}
\end{equation}
\begin{equation}
  V_{\text{gauge}}  =\frac{3 m_X^2}{2 }\left(\frac{a_d}{a}\right)^3 \int\frac{d^3 \vec p}{(2\pi)^3} \frac{ f( |\vec p|,a_d)}{\sqrt{ m_X^2+\left(\frac{a_d}{a}\right)^2|\vec p|^2}} \, ,
\end{equation}
where $a_d$ is the scale factor at the decoupling. We assume that the dark gauge boson is relativistic at $a_d$. The peak of $f(|\vec{p}|,a_d)$ is at $|\vec{p}| \sim T_d$, hence $|\vec{p}|a_d/a \sim T_f =T_d a_d/a$.
 If $m_X^{} \gg T_f$, the $m_X^{}$ in the square root gives the dominating contribution, so the $V_{\text{gauge}} \sim m_X^{} n^{}_X \sim \rho_X^{}\propto m_X^{}/a^3$. On the other hand, if $m_X^{} \ll T_f$, the $m_X^{}$ inside the 
square root of Eq.~\eqref{eq:Vgauge_form} can be neglected resulting in $V_{\text{gauge}}\sim   \rho_X^{} m_X^2/T_f^2\propto m_X^2/a^2 $, where we approximate $\rho_X$ as follows:
\begin{equation}
 \rho_X^{}= 3\left(\frac{a_d}{ a}\right)^3 \int\frac{d^3 \vec p}{(2\pi)^3}\sqrt{m_X^2+\left(\frac{a_d}{a}\right)^2|\vec p|^2}  f( |\vec p|,a_d) \sim T_f^4 \, .
\end{equation}
Therefore, the $V_{\text{gauge}}$ is suppressed by $m_X^2/T_f^2$ from the $\rho_X$ in the $m_X^{} \ll T_f$ limit. One can numerically check that the given relation between the $V_{\text{gauge}}$ and $\rho_X$ is satisfied by the thermal distribution with high accuracy.

\subsection{Thermally coupled}\label{subsec:Thermally coupled}
In this case, the integral in Eq.~\eqref{eq:Vgauge_1} can be evaluated in both the relativistic and non-relativistic limit with the Bose-Einstein distribution
\begin{equation}
    f(|\vec p|,a)= \frac{1}{e^{\sqrt{m_X^2+|\vec p|^2}/T(a)}-1} \, .
\end{equation}

In the relativistic limit,
\begin{equation}
    \langle X_{\mu}X^{\mu} \rangle = \frac{T(a)^2}{4} \, ,
\end{equation}
thus the $V_{\text{gauge}}$ is
\begin{equation}
    V_{\text{gauge}} \approx \frac{ T(a)^2 m_X^2}{8} \, .
\end{equation}
In the non-relativistic limit, 
\begin{equation}
   \langle X_{\mu}X^{\mu} \rangle = \frac{3 T(a)^{3/2}}{(2\pi)^{3/2}}m_X^{1/2} e^{-m_X^{}/T(a)} \, ,
\end{equation}
thus the $V_{\text{gauge}}$ is
\begin{equation}
V_{\text{gauge}}
\approx \frac{3 T(a)^{3/2}}{2(2\pi)^{3/2}}m_X^{5/2} e^{-m_X^{}/T(a)} \, .
\end{equation}
Thus, the $V_{\text{gauge}}$ is exponentially suppressed when $m_X^{} \gg T(a)$.

\subsection{Boltzmann equations}\label{sec:boltz equations}
In this appendix, we derive the Boltzmann equation for mass varying particles. As we will demonstrate, the mass varying effect can be understood as an energy exchange between the quintessence scalar and the dark gauge boson. We start from the general form of the Boltzmann equation of the phase space density $f$.
\begin{equation}
 \mathbf{L}[f] = \mathbf{C}[f] \, ,
\end{equation}
where $\mathbf{L}[f]$ is the Liouville operator and the $\mathbf{C}[f]$ is the collision operator. The covariant and relativistic Liouville operator is  \cite{Kolb:1990vq}
\begin{equation}
    \mathbf{L} =p^{\alpha}\frac{\partial}{\partial x^{\alpha}} -\Gamma^{\alpha}_{\beta\gamma}p^{\beta}p^{\gamma}\frac{\partial}{\partial p^{\alpha}} \, .
\end{equation}

In the FLRW metric, the Boltzmann equation is written as
\begin{equation}
    E\frac{\partial f}{\partial t} -\frac{\dot{a}}{a} E |\vec{p}| \frac{\partial f}{\partial |\vec{p}|} = C[f] \, ,
\end{equation}
where $E=\sqrt{m^2+|\vec{p}|^2}$. If one integrates this equation by the three-momentum, with the integration by part, one gets
\begin{equation}
\begin{split}
   & \frac{d}{dt}\int\frac{d^3\vec p}{(2\pi)^3} E f  - \int\frac{d^3\vec p}{(2\pi)^3} \frac{\partial E}{\partial t}f- \int\frac{d|\vec p|}{2\pi^2}\frac{\partial }{\partial|\vec{p}|}\left(\frac{\dot{a}}{a} E |\vec p|^3f \right)  +  \frac{\dot{a}}{a}\int\frac{d|\vec p|}{2\pi^2} \left(3|\vec{p}|^2E+\frac{|\vec p|^4}{E}\right)f \\&=\int\frac{d^3 \vec p}{(2\pi)^3}C[f] \, .
\end{split}
\label{eq:a1following}
\end{equation}
Since the integral of $f$ in the whole momentum space is finite, $f$ should be 0 in the $|\vec{p}| \rightarrow \infty$ limit.
Also, it is trivial that $|\vec{p}|^3= 0$ at $|\vec{p}|=0$.
Hence, the third term, which is a boundary term, in the l.h.s. should vanish, and the l.h.s can be simplified as
\begin{equation}
    \begin{split}
   & \frac{d}{dt}\int\frac{d^3\vec p}{(2\pi)^3} E f  - \frac{\dot{m}}{m} \int\frac{d^3\vec p}{(2\pi)^3} \frac{(E^2-p^2)}{E}f+  3\frac{\dot{a}}{a}\int\frac{d^3\vec p}{(2\pi)^3} \left(E+\frac{|\vec p|^2}{3E}\right)f  \, ,
\end{split}
\label{eq:a1following2}
\end{equation}
where we used that $\frac{\partial E}{\partial t} = \frac{\dot{m}}{m}\frac{(E^2-p^2)}{E}$. 

The expression can be further simplified with the definitions for the energy density ($\rho$) and the pressure ($p$),
\begin{equation}
    \begin{split}
        \rho&= g\int\frac{d^3\vec p}{(2\pi)^3} E f \, , \\
        p&= g \int\frac{d^3\vec p}{(2\pi)^3} \frac{|\vec p|^2}{3E}   f \, ,
    \end{split}
    \label{eq:rhoandp}
\end{equation}
where $g$ is a degree of freedom of the species. Now, Eq.~\eqref{eq:a1following} can be written as
\begin{equation}
\begin{split}
\dot{\rho} +3H(\rho+p)= \frac{\dot{m}}{m}(\rho-3 p) +g\int\frac{d^3\vec p}{(2\pi)^3}C[f] \, . 
\end{split}
\label{eq:final_boltz}
\end{equation}
The first term in the r.h.s. accounts for the mass varying effect. The second term accounts for the decay and the annihilation of the particle. This equation fully explains the evolution of the energy density of the mass varying particle in the presence of the interactions. When there are no interactions, i.e., $C[f] = 0$, we can see our equation rederives the equation obtained in other mass varying particle scenarios without interactions (for instance, the mass varying neutrino model \cite{Brookfield:2005bz}).

One can show that the $\dot{m}$ dependent term in Eq.~\eqref{eq:final_boltz} is simply a manifestation of the energy conservation between the quintessence scalar and the dark gauge boson. Let us consider a situation that both the quintessence and dark gauge boson do not have any collision terms. The equation of motion of the quintessence scalar can be written in the form of the Boltzmann equation, if one uses the relations $\rho_{\phi}=\dot{\phi}^2/2+V(\phi)$, $p_{\phi}=\dot{\phi}^2/2-V(\phi)$ where $V(\phi) \equiv V_{\text{eff}}-g_X^2\langle X_{\mu}X^{\mu}\rangle\phi^2/2$ as
\begin{equation}
    \dot{\rho}_{\phi}+3H(\rho_{\phi}+p_{\phi})+g_X^2\langle X_{\mu}X^{\mu}\rangle\phi\dot{\phi} = 0 \, .
\end{equation}
The last term can be rewritten with help of Eqs.~\eqref{eq:<XX>} and \eqref{eq:rhoandp} along with $m_X^{} \approx g_X\phi$ as
\begin{equation}
    g_X^2\langle X_{\mu}X^{\mu}\rangle\phi\dot{\phi}  = \frac{\dot{m}_X}{m_X^{}}\int \frac{d^3\vec p}{(2\pi)^3} \frac{3 m_X^2 f(\vec p)}{E_{\vec p}} = \frac{\dot{m}_X}{m_X^{}} (\rho_X-3p_X) \, .
\end{equation}
Hence, the Boltzmann equations for the dark gauge boson and the quintessence are
\begin{equation}
\begin{split}
    & \dot{\rho}_{\phi}+3H(\rho_{\phi}+p_{\phi})=-\frac{\dot{m}_X}{m_X^{}} (\rho_X-3p_X) \, , \\ &\dot{\rho}_{X}+3H(\rho_{X}+p_{X})=\frac{\dot{m}_X}{m_X^{}} (\rho_X-3p_X) \, .
\end{split}
\label{eq:finalboltz}
\end{equation}
The r.h.s. of these equations tell us that the energy transfer between the quintessence scalar and the dark gauge boson is proportional to the $\dot{m}_X$.

\section{Coherent dark gauge boson}\label{sec:Ultra-light dark gauge boson}
If the dark gauge boson is in a homogeneous condensate state, the equations of motion in Eq.~\eqref{eq:eom} become \cite{Nelson:2011sf}
\begin{equation}
\begin{split}  
\ddot{\phi} + 3 H \dot{\phi}+\frac{\partial V_0}{\partial \phi}+g_X^2\frac{|\vec{X}|^2}{a^2} \phi&=0 \, , \\
\ddot{\vec{X}}+H \dot{\vec{X}}+g_X^2\phi^2\vec{X} = 0 \, ,
\end{split}
\label{eq:condensateeom}
\end{equation}
and the energy density of the dark gauge boson is 
\begin{equation}
    \rho_X^{}=\frac{1}{2a^2} \big( |\dot{\vec{X}}|^2+g_X^2\phi^2|\vec{X}|^2 \big) \, .
    \label{eq:condensatedensity}
\end{equation}
Note that $g_X\phi \approx m_X^{}$ and $X_{0}=0$ \cite{PhysRevD.100.063541,Nakayama:2019rhg}.

(i) In the early universe, $H\gg m_X^{}$ and the dark gauge boson field stay frozen due to the friction term. One finds from Eq.~\eqref{eq:condensateeom} that
\begin{equation}
     \frac{1}{a} \frac{d}{dt}(a \dot{\vec{X}}) \approx 0 \, .
\end{equation}
So,
\begin{equation}
    \vec{X}= \vec{X}_a + \vec{X}_b \int \frac{dt}{a} \, ,
\end{equation}
where $\vec{X}_a$ is the initial $\vec{X}$ value, and $\vec{X}_b$ is some proportionality constant. 

Thus, if one assumes the initial condition that $\dot{\vec{X}} \approx 0$, i.e., $ |\vec{X}_b| \ll |\vec{X}_a|$, the energy density of the dark gauge boson evolves as $\rho_X^{}\propto g_X^2\phi^2 a^{-2} \approx m_X^2 a^{-2}$ since the $\vec{X}$ is almost constant. Also, 
\begin{equation}
    V_{\text{gauge}} = \frac{1}{2a^2}g_X^2\phi^2|\vec{X}|^2 \approx \rho_X^{}\, .
\end{equation}

(ii) As the universe ages, the $H$ may drop below the $m_X^{}$. Then the $\vec{X}$ oscillates as Eq.~\eqref{eq:condensateeom} suggests. The dark gauge boson equation of motion in Eq.~\eqref{eq:condensateeom} can be simplified with the conformal coordinate ($\tau,\vec{x}$) where $d\tau=dt/a$ as
\begin{equation}
\frac{d^2X_i}{d\tau^2}+a^2 m_{X}^2X_i=0 \, ,
\label{eq : simcondensateq}
\end{equation}
and each $X_i$ is independent of the other components. 
One can use the WKB approximation to find an approximated solution when the following condition is satisfied \cite{BAND2013303}
\begin{equation}
 \dfrac{d(a m_X^{})}{d\tau} \ll a^2m_X^2 \, .
 \label{eq:wkb}
\end{equation}
This condition is equivalent to the adiabatic condition \eqref{eq:adiabaticity}.
Then the WKB-like solution of Eq.~\eqref{eq : simcondensateq} \cite{BAND2013303} is
\begin{equation}
 X_i(\tau) \approx \text{Re} \left[\frac{\chi^{}_i}{\sqrt{a m_X^{}}} e^{i \int d\tau\, a m_X^{}} \right] \, ,
 \label{eq:wkbXi}
\end{equation}
where $\text{Re}[f]$ denotes the real part of $f$ and $\chi_i$ are some coefficient. We can take $\vec{X}(\tau) =(0,0,X_3(\tau)) $ by choosing a proper coordinate without losing generality \cite{Nakayama:2019rhg}.

Also one can show that the $\dot{X_3}$ is oscillating a quarter cycle away from the $X_3$,
\begin{equation}
\begin{split}
    \dot{X_3}(\tau)& \approx \text{Re}\left[\left( -\frac{1}{2a^2}\frac{d a}{d\tau}-\frac{1}{2am_X^{}}\frac{d m_X^{}}{d\tau} +i m_X^{}  \right)X_3(\tau)\right] \\
    &\approx \text{Re}\left[\frac{m_X^{} \chi^{}_3}{\sqrt{a m_X^{}}} e^{i(\pi/2 + \int d\tau\, a m_X^{})} \right] \, ,
\end{split}
\end{equation}
where the condition~\eqref{eq:wkb} is used in the second line. Therefore the energy density of the dark gauge boson is
\begin{equation}
    \rho_X^{}\approx \frac{ m^{}_X }{2a^3}|\chi_3|^2 \propto \frac{m_X^{}}{a^3} \, .
\end{equation}
Note that the magnitude of the $|\dot{\vec{X}}|^2$ and the $m^2 |\vec{X}|^2$ is the same. Therefore, we have 
\begin{equation}
    V_{\text{gauge}} \approx \frac{1}{2}\rho_X^{}\,.
    \label{eq:vgaugecoherentosc}
\end{equation}

\bibliographystyle{JHEP}
\bibliography{Nwriteup.bib}

\end{document}